\documentclass[a4,letters,usenatbib]{mnras}
\usepackage{amsmath}
\usepackage{amssymb} 
\usepackage{multirow}
\usepackage{textcomp}
\usepackage{microtype}
\usepackage{subcaption}
\usepackage[dvipsnames]{xcolor}
\usepackage{ulem}
\usepackage{graphicx}
\usepackage{float}
\usepackage{mathptmx}
\usepackage{tabularx}
\usepackage{caption}
\usepackage{acronym}
\usepackage{ulem}
\usepackage{bm}

\newacro{PTA}{pulsar timing array}
\newcommand{\PTA}{\ac{PTA}}
\newacro{SNR}{signal-to-noise ratio}
\newcommand{\SNR}{\ac{SNR}}
\newacro{GW}{gravitational wave}
\newcommand{\GW}{\ac{GW}}

\newcommand{\be}{\begin{equation}}
\newcommand{\ee}{\end{equation}}
\newcommand{\bdm}{\begin{displaymath}}
\newcommand{\edm}{\end{displaymath}}
\newcommand{\bea}{\begin{eqnarray}}
\newcommand{\eea}{\end{eqnarray}}

\newcommand{\Mchirp}{\mathcal{M}}

\newcommand{\loc}{{\Omega}_{90}}
\newcommand{\hp}{h^{+}}
\newcommand{\hc}{h^{\times}}
\newcommand{\Hp}{H^{+}}
\newcommand{\Hc}{H^{\times}}
\newcommand{\Hvec}{\bm{H}}

\newcommand{\Dspace}{\mathbb{D}}
\newcommand{\Nullspace}{\mathbb{A}}
\newcommand{\dvec}{\bm{d}}
\newcommand{\HpF}{\tilde{H}^{+}}
\newcommand{\HcF}{\tilde{H}^{\times}}
\newcommand{\Ap}{A^{+}}
\newcommand{\Ac}{A^{\times}}

\newcommand{\Fp}{F^{+}}
\newcommand{\Fc}{F^{\times}}
\newcommand{\Fpv}{\bm{F}^+}
\newcommand{\Fcv}{\bm{F}^{\times}}
\newcommand{\Fmat}{\bm{\mathsf{F}}}
\newcommand{\FMP}{\bm{\mathsf{F}}^{-1}_{MP}}
\newcommand{\Mmat}{\bm{\mathsf{M}}}
\newcommand{\Amat}{\bm{\mathsf{A}}}
\newcommand{\Smat}{\bm{\mathsf{S}}}
\newcommand{\Pmat}{\bm{\mathsf{P}}}
\newcommand{\Imat}{\bm{\mathsf{I}}}
\newcommand{\Qmat}{\bm{\mathsf{Q}}}
\newcommand{\Rmat}{\bm{\mathsf{R}}}
\newcommand{\Gammamat}{\bm{\mathsf{\Gamma}}}
\newcommand{\params}{\bm{\lambda}}
\newcommand{\Dt}{\Delta t}

\def\lsim{\lower.5ex\hbox{$\; \buildrel < \over \sim \;$}}
\def\msun{\,{\rm M_\odot}}

\title[Null Stream localisation of GW sources in PTAs]
{\textbf{Null stream analysis of Pulsar Timing Array data: localisation of resolvable gravitational wave sources}}

\author[Goldstein et al.]
       {Janna Goldstein$^{1}$\thanks{E-mail: jgoldstein@star.sr.bham.ac.uk}, John Veitch$^{1,2}$, Alberto Sesana$^{1}$ \& Alberto Vecchio$^{1}$
\\ %
$^{1}$ School of Physics and Astronomy and Institute for Gravitational Wave Astronomy, University of Birmingham, Edgbaston B15 2TT, United Kingdom\\
$^{2}$ Institute for Gravitational Research, University of Glasgow, Glasgow, G12 8QQ, United Kingdom}

\begin{document}

\date{}

\pagerange{\pageref{firstpage}--\pageref{lastpage}} \pubyear{2017}

\maketitle

\label{firstpage}

\begin{abstract}
		Super-massive black hole binaries are expected to produce a \GW\ 
		signal in the nano-Hertz frequency band which may be detected by \acp{PTA} in the coming years. The signal is composed of both stochastic and individually resolvable
		components.
		Here we develop a generic Bayesian method for the analysis of resolvable sources based on the construction of `null-streams' which cancel the part of the signal held in common for each pulsar (the Earth-term). For an array of $N$ pulsars there are $N-2$
		independent null-streams that cancel the \GW\ signal
		from a particular sky location. This method is applied to the localisation of quasi-circular binaries undergoing adiabatic inspiral. We carry out a systematic investigation of the scaling of
		the localisation accuracy with signal strength and number of pulsars in
		the \PTA. Additionally, we find that source sky localisation with the
		International PTA data release one is vastly superior
		than what is achieved by its constituent regional \PTA s.

\end{abstract}
\acresetall

\section{Introduction}
The recent successes of the LIGO-Virgo collaboration \citep{2016PhRvL.116f1102A,2016PhRvL.116x1103A,2017PhRvL.118v1101A} brought \GW\  astronomy in the spotlight. Despite their great achievements, ground based interferometers are only sensitive in a frequency range from about 10 Hz to 1000 Hz and are thus suited for detection of stellar mass compact objects such as stellar mass black holes or neutron stars \citep{2010CQGra..27q3001A}. The \GW\  spectrum however, extends for several more decades in frequency \citep{2017ogw..book...43C}. In particular, the low frequency band, is expected to be dominated by \GW\  signals coming from a class of much more massive astrophysical sources: supermassive black hole binaries \citep[SMBHBs,][]{1980Natur.287..307B}.

The adiabatic inspiral of $10^8-10^{10} \msun$ SMBHBs at cosmological distances generates loud \acp{GW} in the nHz-to-$\mu$Hz frequency range \citep[see, e.g.,][]{2008MNRAS.390..192S}, where ground based interferometers are completely deaf. Fortunately, Nature provided us with formidably stable natural clocks that might allow to hear such low frequency waves in the foreseeable future: millisecond pulsars \citep[MSPs,][]{2008LRR....11....8L}. Located at kpc distances within the Galaxy, MSPs behave like cosmic lighthouses sending periodic radio signals to the Earth. If a \GW\ crosses the path of the radio photons, their null geodesic is modified, effectively resulting in a \GW\ induced redshift \citep{1978SvA....22...36S}. In practice radio pulses arrive on Earth a little bit earlier or later than expected, an effect that can be measured if the time of arrivals (TOAs) of the radio pulses can be determined with enough precision. The TOAs of the most stable MSPs can be currently determined with an uncertainty of about 100ns \citep{IPTADR1:2016}, an accuracy level approaching the expected delays induced by the most massive SMBHBs populating the Universe \citep{2009MNRAS.394.2255S}.

SMBHBs are expected to be common in the Universe, and \acp{PTA} will be mostly sensitive to the incoherent superposition of \acp{GW} coming from the large population of these cosmological sources \citep{1995ApJ...446..543R,2003ApJ...583..616J,2008MNRAS.390..192S}. At the high mass end and for sufficiently high frequencies however, SMBHBs become sparser, and the loudest ones will likely be individually detectable as deterministic sources \citep{2009MNRAS.394.2255S}. Consequently, several algorithms and pipelines have been assembled in recent years to detect and characterize both a stochastic \GW\ background and individual deterministic sources \citep[see][for a recent review]{2017arXiv170902816P}. In both cases the challenge is to determine whether the data are better described by noise only or noise plus some \GW\ signal. No \GW\ detection has been reported thus far, and several pipelines have been used to produce upper limits on the strength of each type of source \citep{2014ApJ...794..141A,2014MNRAS.444.3709Z,2015Sci...349.1522S,2015MNRAS.453.2576L,2015PhRvL.115d1101T,2016MNRAS.455.1665B,2016ApJ...821...13A}.

The problem of detecting a \GW\ signal in \PTA\ data is complicated by the variety of noise sources, that can be either peculiar to each pulsars (e.g., spin modifications due to movements in the pulsar crust) or common in all observed systems (e.g., an error in the time standard used as a reference to measure TOAs).
The latter are more insidious as they may introduce correlations in the residuals between the timing model and the observed times of arrival.
The Hellings and Downs curve describes the cross-correlation due to an isotropic GW background between pairs of pulsars as a function of their angular separation on the sky~\citep{hellings83}.
Common noise sources such as errors in the Solar System ephemeris or
clock errors also produce cross-correlations between
pulsars which may be confused with a \GW\  signal \citep{Tiburzi:2016MNRAS, Taylor:2017PhRvD}.

Previous approaches to the detection problem have been developed which
marginalise over these errors by including uncertainties in the timing
model itself. However, these methods are still vulnerable to unmodelled
systematics which may remain. For robust detection one would like to
have an empirical estimate of the noise with which to compare the
observed data. It is difficult to produce such an empirical
background distribution due to the finite
amount of \PTA\ data available and the fact that the \GW\ signal cannot be removed from the data.
This issue has been addressed by the development of techniques to \textit{e.g.} scramble the timing residuals 
so that the GW-induced cross-correlations are
not present in the scrambled data realisation~\citep{Taylor:2017PhRvD}.

In this work we investigate an alternative approach applicable to deterministic and individually resolvable signals.
Through this method one cancels the gravitational wave signal exactly by exploiting redundancies in the data when the number of data streams exceeds
the number of independent degrees of freedom  (i.e. the number of polarizations) in the GW signal. Given independent data streams from $N$ detectors and $M$ GW polarizations, one can construct $M$ GW polarization streams and $N-M$ 'null streams', which have all  \GW\ power from a particular direction removed. This allows a discrimination between \acp{GW} and noise which can be used to construct a statistical model of the data. Assuming as working hypothesis that General Relativity holds, only the two tensor polarizations of the GW field are non-vanishing, thus allowing the construction of $N-2$ null streams.

The null-stream formalism is quite general and has been applied to
analyses across the gravitational-wave spectrum.
For networks of ground-based detectors the method has been proporsed
to discriminate between signal and unmodelled
noise~\citep{2005CQGra..22S1321W,2006CQGra..23S.741A,2006PhRvD..74h2005C,2006CQGra..23S.673R}.
In the context of the Laser Interferometer Space Antenna (LISA)
an example of null stream is given by the Sagnac configuration of
the detector~\citep{2017arXiv170200786A}, in which the interferometer
channel are combined to cancel out \GW\ signals, thus serving as
detector calibrator to assess the instrumental noise
level~\citep{2004PhRvD..69b2001S}. Recently,
\citet{2015MNRAS.449.1650Z,2016MNRAS.461.1317Z}
adapted those techniques to PTAs, and investigated the benefits
of using null streams to reconstruct the \GW\ signal properties
and quantify detection confidence in a frequentist framework. 

Here we develop a Bayesian \PTA\ analysis using the null-stream formalism for an arbitrary deterministic \GW\ source. We derive the associated
likelihood function and use it for the recovery of the 
source properties.
Although the null-stream formalism works in both the time and frequency
domains given appropriate interpolation, 
for simplicity we consider the frequency domain analysis of simulated data
containing a monochromatic \GW\ source from a
supermassive black hole binary \citep{2016arXiv160703459H}. 

As a first application of the method, we investigate its performance in localising resolvable SMBHBs. Sky localisation is of paramount importance for PTA science, because it opens the possibility of identifying the source galactic host and of looking for possible electromagnetic counterparts; consequently it has been tackled by serveral authors in recent years \citep{2010PhRvD..81j4008S,Lee:2011MNRAS,2012PhRvD..85d4034B,2012ApJ...756..175E,2014PhRvD..90j4028T,2015MNRAS.449.1650Z,2016MNRAS.461.1317Z,2017PhRvL.118o1104W}. For this specific problem, the null stream techinque is expected to produce equivalent results to other analysis methods (exploiting a linear transformation on the data). However, this application serves to lay out the formalism in the Bayesian framework. This will be used in future work to exploit the main advantage of null streams: by creating combinations of data that contain noise only, they are a powerful tool to discriminate signal from noise, thus allowing to tackle the issue of detection confidence, which is critical in PTA data analysis \citep{Taylor:2017PhRvD}. We perform a systematic investigation of the source sky localisation as a function of \SNR\ and number of pulsars in the array. We then consider the case of a detection with current PTAs, demonstrating the great benefits of combining regional \PTA\ data under the aegis of IPTA.

The paper is organized as follows. In Sec. \ref{sec:method} we describe the
null-stream construction and the Bayesian framework employed to extract source
properties from the data. In Sec. \ref{sec:results} we investigate systematically the scaling of source sky localisation capabilities as a function
of the main \PTA\ parameters and compare our results with previous work based on
the Fisher Matrix approximation. In Sec. \ref{sec:implications} we apply our
techniques to current PTAs and demonstrate the benefit of the world-wide IPTA
network. We summarize our main findings and discuss future prospects for
expanding this work in Sec. \ref{sec:conclusions}.

\section{Method}
\label{sec:method}
The basic idea behind the null stream method is the following: data is obtained 
from $N$ detectors that have a linear response to a \GW\ signal. The two 
polarizations of the \GW\ can be reconstructed from the detector output, which 
leaves the possibility to construct an additional $N\!-\!2$ independent data 
combinations. If the detector responses are known, these combinations can be 
made such that any \GW\ signal is cancelled out, leaving only noise, hence the 
name null streams.

The following sections will explain in detail the construction of null streams for Pulsar Timing Array (PTA) data (Sec. \ref{sec:construction}) and our choice of a continuous wave signal (Sec. \ref{sec:continuous_wave}). Then, the use of null streams in the sky localisation of a single source \GW\ signal (Sec. \ref{sec:localisation}) and the implementation for discrete data (Sec. \ref{sec:discrete_data}) are discussed. First in Sec. \ref{sec:PTA_response}, the formalism is set up in terms of the signal and the response of the PTA.

\subsection{Response of a PTA} \label{sec:PTA_response}

Assume we have a plane wave propagating in the direction $\bm{\hat{\varOmega}}$, with angular frequency $\omega$. A coordinate system can be chosen by using $\bm{\hat{\varOmega}}$ and two additional orthonormal vectors:
\begin{align}
	\bm{\hat{\varOmega}} &= \big( \sin\theta \cos\phi,~ \sin\theta \sin\phi,~\cos\theta \big) \nonumber \\
	\bm{\hat{m}} &= \big( \sin\phi,~-\cos\phi,~0 \big) \nonumber \\
	\bm{\hat{n}} &= \big( \cos\theta \cos\phi,~\cos\theta\sin\phi,~-\sin\theta \big).
\end{align}
Here, $\theta$ and $\phi$ are the polar sky coordinates of the direction of propagation of the \GW\ ($-\bm{\hat{\varOmega}}$ points towards the source).
The two orthogonal polarizations of the gravitational wave can be written in terms of the $+$ and $\times$ transverse basis tensors
\begin{align}
	e^+_{ij}(\bm{\hat{\varOmega}}) &= \hat{m}_i\hat{m}_j - \hat{n}_i\hat{n}_j \nonumber \\
	e^{\times}_{ij}(\bm{\hat{\varOmega}}) &= \hat{m}_i\hat{n}_j + \hat{n}_i\hat{m}_j.
\end{align}
The metric perturbation due to the \GW\ then, is given by:
\be
h_{ij}(t) = \hp(t)e^{+}_{ij} + \hc(t)e^{\times}_{ij},
\ee
where $\hp(t)$ and $\hc(t)$ are the amplitudes of the two polarizations.

A \GW\ propagating through the Galaxy affects the travel time of radio emission 
travelling from a pulsar to the Earth. The resulting redshift in the pulse TOAs 
depends on the relative angle between the pulsar, in direction $\bm{\hat{p}}$, and 
the \GW\ propagation direction $\bm{\hat{\varOmega}}$ \citep[see, 
e.g.,][]{Anholm:2008wy, Lee:2011MNRAS}.
\be
z(t, \bm{\hat{\varOmega}}) = \frac{1}{2} \frac{\hat{p}^i\hat{p}^j}{1 + \bm{\hat{\varOmega}} \cdot \bm{\hat{p}}} \Delta h_{ij},
\ee
where $\Delta h_{ij} = h_{ij}(t_{E}, \bm{\hat{\varOmega}}) - h_{ij}(t_{p}, \bm{\hat{\varOmega}})$, being $ h_{ij}(t_{p})$ and $h_{ij}(t_{E})$ the metric perturbation at the pulsar at the time of pulse emission and at the Earth at the time of pulse detection, respectively.
Only the Earth term adds up coherently in the analysis of multiple pulsars in the array.
Therefore, for any burst-like signal with duration shorter than the travel time of the pulses only the Earth term is relevant. For a continuous wave (e.g. from a SMBHB) on the other hand, the pulsar term is present, although its frequency may differ slightly from that of the Earth term as it samples different periods in the wave-train of the slow inspiral \citep{2016MNRAS.455.1665B}. Whether the pulsar and the Earth terms fall at different frequencies or not depends on the intrinsic properties of the GW source, the distance to the pulsar and the relative pulsar-source angular separation. Implementing realistic SMBHB population models and considering plausible developments of current PTAs, \cite{2015MNRAS.451.2417R} found that either situation is possible, with comparable probability. To simplify the problem, as a first implementation, only the Earth term will be considered in our description of the signal \footnote{The addition of the pulsar term is, on the other hand, necessary for non evolving sources, and it might also help improving sky localization, as shown by \cite{Lee:2011MNRAS} and \cite{2016MNRAS.461.1317Z}. We will address this case in future work.}. This results in the following definition for the response functions $\Fp$ and $\Fc$:
\begin{align}
	z(t, \boldsymbol{\hat{\varOmega}}) &= \frac{1}{2} \frac{\hat{p}^i\hat{p}^j} {1 + \boldsymbol{\hat{\varOmega}} \cdot \boldsymbol{\hat{p}}} \big( \hp(t)e^+_{ij} (\boldsymbol{\hat{\varOmega}}) + \hc(t)e^{\times}_{ij} (\boldsymbol{\hat{\varOmega}}) \big) \nonumber \\
	&\equiv F^+ (\boldsymbol{\hat{\varOmega}}) \hp(t) + F^{\times} (\boldsymbol{\hat{\varOmega}}) \hc(t)
	\label{eq:response_functions}
\end{align}

The observables for a \PTA\ are not the redshifts, but the residuals $r(t)$ obtained by taking the difference between the predicted and measured TOAs. The relation between the two is simply that the residuals are the integrated redshifts:
\be
r(t) = \int_0^t z(\tau) d\tau.
\ee
Since the response functions are time independent, applying the integral to the previous Eq. \ref{eq:response_functions} yields:
\be
r(t, \boldsymbol{\hat{\varOmega}}) = F^+(\boldsymbol{\hat{\varOmega}})\Hp(t) + F^{\times}(\boldsymbol{\hat{\varOmega}})\Hc(t),
\label{eq:response_residuals}
\ee
where $\Hp = \int_0^t \hp(\tau) d\tau$ and similarly for $\Hc$. Our final calculations will be done in the frequency domain, for which we can substitute $r \to \tilde{r}$ and $\Hp \to \HpF$, and $\Hc \to \HcF$, since the Fourier transform is linear.

Eq.~\ref{eq:response_residuals} can be written for each pulsar in the PTA, resulting in a collection of residuals $\{r_i\}$. Labelling the response functions $\Fp_i$ and $\Fc_i$ for the pulsar in the direction $\bm{\hat{p}}_i$, this can be combined into the matrix equation:
\be
\boldsymbol{r} = \begin{pmatrix}
	F^{+}_1 & F^{\times}_1 \\
	F^{+}_2 & F^{\times}_2 \\
	\vdots & \vdots \\
	F^{+}_N & F^{\times}_N \\
\end{pmatrix} 
\begin{pmatrix}
	\Hp \\
	\Hc
\end{pmatrix} 
\equiv \Fmat \begin{pmatrix}
	\Hp \\
	\Hc
\end{pmatrix},
\label{eq:response_matrix}
\ee
where we have defined the response matrix $\Fmat$. $\Fmat$ depends on 
the location of the \GW\ source $-\bm{\hat{\varOmega}}$, but not on the parameters 
of the specific form of the \GW\ polarizations, which makes the 
following null stream construction general.

\subsection{Null stream construction} \label{sec:construction}

For a fixed direction $\bm{\hat{\varOmega}}$, the matrix $\Fmat$ defines a mapping from the two-dimensional space of gravitational waves $\Hvec\equiv(\Hp,\Hc)\in \mathbb{R}^2$ \footnote{We drop the dependence on time as the logic applies to any particular time stamp (or Fourier frequency bin when using $\bm{\tilde{H}}$).} to the $N$-dimensional space of residuals from $N$ pulsars in the array. The image of this mapping contains the residuals induced by a gravitational wave, but the measured response data $\bm{d}\in\Dspace$ is subject to additional noise $\bm{n}$ such that $\bm{d}=\Fmat\Hvec + \bm{n}$.
For $N>2$, the space $\Dspace$ can be split into a two-dimensional subspace with the image of $\Fmat$, and an orthogonal $N-2$-dimensional subspace. This second subspace -- the nullspace $\Nullspace$ -- is spanned by a set of $N-2$ independent null-streams, that are also independent of the gravitational wave.

There are different ways to choose the $N\!-\! 2$ 
independent null streams from detectors' output $\bm{d}$ (e.g. 
\cite{2015MNRAS.449.1650Z,2016MNRAS.461.1317Z} use a different method than us). 
However, it is possible to impose the more stringent requirement that the null 
streams are orthogonal. The method that we describe here has been adapted from 
work by \citet{Chatterji:2006} and \citet{Rakhmanov:2006qm}. In short, it's a construction of a set of orthogonal basis vectors for the nullspace $\Nullspace$.

We are interested in inverting the mapping $\Fmat$ from gravitational-waves to 
a given set of residual data so that we may reconstruct the signal and find the 
null-streams. The maximum likelihood solution 
for $\Hvec$ is given by $\hat\Hvec = \FMP\dvec$,
where $\FMP \equiv (\Fmat^{\top} \Fmat)^{-1} \Fmat^{\top}$ is the Moore-Penrose 
pseudo-inverse of $\Fmat$~\citep{Rakhmanov:2006qm}.
This can be seen as a projection of the data onto the two dimensional subspace of $\Dspace$ spanned by the column vectors $\Fpv$ and $\Fcv$ of the response matrix $\Fmat$. The null streams are found by projecting onto the orthogonal space $\Nullspace$, to ensure the null-streams contain no component of the signal. Say a 
basis for the null space is $\{\bm{\hat{e}}_i\}$, with $\bm{\hat{e}}_i \cdot \Fpv = 
\bm{\hat{e}}_i \cdot \Fcv = 0$ where $i \in 1, ..., N\!-\!2$ labels the basis vectors. Then the matrix $\Amat$ 
with rows $\bm{\hat{e}}_i^{\top}$ is the nullspace projection matrix because 
$\Amat \Fmat = \bm{\mathsf{0}}$ as per construction (where $\bm{\mathsf{0}}$ is a $(N-2)\times2$ zero matrix). The $N\!-\!2$ null streams can 
then be calculated as:
\be \Amat \bm{d} = \Amat \big( \Fmat \begin{pmatrix}
H^{+} \\
H^{\times}
\end{pmatrix} + \bm{n} \big) = \bm{\eta} + \Amat \bm{n}.
\ee
Here we define $\bm{\eta}$ to be the vector of null streams which all equal zero ($\eta_i = 0$).

To find the basis $\{\bm{\hat{e}}_i\}$, consider the projection operators 
\mbox{$\Pmat = \Fmat \FMP$}, and $\Smat = \Imat - \Pmat$, where 
$\Imat$ is the $(N\! \times\! 
N)$ identity matrix~
\citep[see also][]{Rakhmanov:2006qm}. The first projects onto the column space:
\be
\Pmat \Fmat = \Fmat \FMP \Fmat = \Fmat,
\ee
and the second onto the null space:
\be
\Smat \Fmat = (\Imat - \Pmat) \Fmat = \Fmat - \Fmat = 0.
\ee
However, $\Smat$ is an $(N\! \times\! N)$ matrix whereas the null space only has $N\!-\!2$ dimensions. A way to reduce $\Smat$ to $((N\!-\!2)\! \times N)$ is to use the QR-decomposition, which yields $\Smat = \Qmat \Rmat$. Then, if $\Smat$ has rank $r$, the first $r$ columns of $\Qmat$ form an orthonormal basis for the column vectors of 
$\Smat$. Therefore, the first $N\!-\!2$ columns of $\Qmat$ form the basis $\{\bm{\hat{e}}_i\}$ that we were looking for.

Because both the reconstructed \GW\ polarizations and the null streams are 
informative, the projectors $\FMP$ and $\Amat$ are combined in the square 
matrix $\Mmat$. The total projection of the data with $\Mmat$ is:
\begin{align}
\Mmat\, \bm{d} &= \begin{pmatrix}
	(\FMP)_1 \\
	(\FMP)_2 \\
	\hat{e}_1 \\
	\vdots \\
	\hat{e}_{N-2}
\end{pmatrix} \bigg( \Fmat \begin{pmatrix}
\Hp \\
\Hc
\end{pmatrix} + \bm{n} \bigg)
= \begin{pmatrix}
	\Hp \\
	\Hc \\
	\eta_1 \\
	\vdots \\
	\eta_{N-2}
\end{pmatrix} + \Mmat\, \bm{n} \\
&\equiv \bm{h} + \Mmat\, \bm{n},
\label{eq:strainspace_data}
\end{align}
where we have defined $\bm{h}$ as the combined vector of \GW\ polarizations and null streams.

\subsection{Continuous wave signal} \label{sec:continuous_wave}

The null stream method can be used with any assumption on the functional form 
of the \GW\ polarizations $\hp(t)$ and $\hc(t)$.
Here we specialise and assume the 
signal to take the form of a monochromatic continuous wave from a circular 
SMBHB. The monochromatic assumption is valid so long as the SMBHB is light
enough such that any frequency evolution over the time scale of the observation 
is negligible, which is generally true for relevant systems 
\citep{2010PhRvD..81j4008S, Taylor:2015kpa}.
Both polarizations $\hp$ and $\hc$ are related, via the angle $\psi$,
to the \GW\ signal emitted by the source:
\begin{align}
	\hp(t) &= \Ap(\params) \cos{(2\psi)} - \Ac(\params) \sin{(2\psi)}
	\label{eq:general_hp} \\
	\hc(t) &= \Ap(\params) \sin{(2\psi)} + \Ac(\params) \cos{(2\psi)},
	\label{eq:general_hc}
\end{align}
where
\begin{align}
\Ap &= A \frac{1}{2} (1 + \cos{\iota}^2) \cos{(\omega_0 t + \phi)} \label{eq:polarization_Ap} \\
\Ac &= A (\cos{\iota}) \sin{(\omega_0 t + \phi)}.
\label{eq:polarization_Ac}
\end{align}
The frequency of the \GW\ is $f_0 = \omega_0/2\pi$ (which is twice the orbital 
frequency). For a chosen frequency and sky position, the remaining parameters 
of the source are the binary's orbital inclination $\iota$, the 
polarization angle $\psi$, the phase offset $\phi$ and the amplitude $A$, which we encapsulate in the parameter vector $\params$. The amplitude depends on the physical parameters of the SMBHB:
\be
A = \frac{4 \Mchirp (\pi f_0)^{2/3}}{D_l},
\ee
where $\Mchirp$ is the redshifted chirp mass of the binary, $D_l$ the luminosity distance to the source and $f_0$ the observed \GW\ frequency (here $G=c=1$). However in this work, we treat $A$ as an overall scaling factor of the signal.

The form of the signal needs to be changed when considering the \PTA\ residuals instead of the redshifts, as in Eq. \ref{eq:response_residuals}. Applying the time integral to Eqs. \ref{eq:polarization_Ap} and \ref{eq:polarization_Ac} yields:
\begin{align}
\Ap_{(t)} &\equiv \int_0^t \Ap(\tau) d\tau = \frac{A}{2\omega_0} (1 + \cos{\iota}^2) \sin{(\omega_0 t + \phi)} \label{eq:int_polarization_Ap}\\
\Ac_{(t)} &\equiv \int_0^t \Ac(\tau) d\tau = -\frac{A}{\omega_0} \cos{\iota} 
\cos{(\omega_0 t + \phi)},
\label{eq:int_polarization_Ac}
\end{align}
where we disregard constants of the integration.

\subsection{Localisation} \label{sec:localisation}

The predictable shape of the null streams (they contain only noise) can have 
many applications. For example, the null stream statistic should follow the 
statistic of the noise and can therefore be used to validate candidate \GW\ 
signals and assess detection confidence, which we plan to investigate in the 
future. In this work, we use it to estimate the sky location of a \GW\ source. 
Only when constructing the response function and the matrix $\Mmat$ using the 
correct sky location, do the signal components in the null streams cancel out. 
Thus, an estimate for the sky location is obtained by varying $\boldsymbol{\hat{\varOmega}}$ 
until the null streams are closest to zero (and, consequently, the \GW\ 
polarizations closely match the model). 

To quantify this, consider the posterior 
distribution on the sky location, under the assumption $\mathcal{H}_{sig}$ that 
a signal is present:
\be
p(\boldsymbol{\hat{\varOmega}} | \bm{d}, \mathcal{H}_{sig}) = \frac{p(\boldsymbol{\hat{\varOmega}} | 
\mathcal{H}_{sig}) p(\bm{d} | \boldsymbol{\hat{\varOmega}}, \mathcal{H}_{sig})}{p(\bm{d})}.
\ee
The prior on the sky location $p(\boldsymbol{\hat{\varOmega}} | \mathcal{H}_{sig})$ is assumed 
to be flat. To calculate the likelihood, a model for the data is needed.
In the presence of 
a signal and additive Gaussian noise, Eq. \ref{eq:strainspace_data} describes 
what is needed: $\bm{d} = 
\Mmat^{-1} \bm{h} + \bm{n}$. This naturally leads to the Gaussian 
log-likelihood function:
\be
l = -\frac{1}{2} \bigg( (\bm{d} - \Mmat^{-1}\bm{h})^{\top} \Gammamat (\bm{d} - 
\Mmat^{-1}\bm{h}) \bigg) + norm.
\label{eq:likelihood_inverse}
\ee
where $\Gamma$ is the inverse of the covariance matrix appropriate for the 
expected noise of the detector. The normalisation is not written explicitly, as 
the likelihoods are normalised numerically as a last step in the calculation.
Eq. \ref{eq:likelihood_inverse} can be rewritten using $\Imat = \Mmat^{-1} \Mmat$ to the following form:
\be
l = -\frac{1}{2} \bigg( (\Mmat\bm{d} - \bm{h})^{\top} ((\Mmat^{-1})^{\top} 
\Gammamat \Mmat^{-1}) (\Mmat\bm{d} - \bm{h}) \bigg) + norm.
\label{eq:likelihood}
\ee
The $\Mmat^{-1}\bm{h}$ term in Eq. \ref{eq:likelihood_inverse} depends on both the sky location $\bm{\Omega}$ through $\Mmat$ and the GW model parameters $\bm{\lambda}$ through $\bm{h}$. In Eq. \ref{eq:likelihood} these depencies are split up over the terms $\Mmat \bm{d}$ and $\bm{h}$, which simplifies calculations. 

To obtain the likelihood $p(\bm{d} | \boldsymbol{\hat{\varOmega}}, \mathcal{H}_{sig})$, this 
$l$ is marginalised over the \GW\ parameters $\bm\lambda$. By having split the dependency on $\bm{\lambda}$ from $\Mmat\bm{d}$, this quantity has to be calculated only once for each sky location. For our choice of a 
continuous wave signal in Section \ref{sec:continuous_wave}, the marginalization is done with 
a combination of an analytical and numerical integration. A benefit to the 
particular method of null stream construction used (Sec. \ref{sec:construction}) 
becomes apparent here. For a diagonal covariance matrix $\Gammamat^{-1}$ of the 
detector noise, the transformed matrix $(\Mmat^{-1})^{\top} \Gammamat \Mmat^{-1}$ 
is largely kept diagonal (except for the covariance between the \GW\ 
polarization amplitudes in the first two entries of $\Mmat\bm{d}$ and of 
$\bm{h}$), which can make the numerical computation more efficient.

To quantify how well a \GW\ source is localised,
we define $\loc$ as the fraction of the sky area 
containing $90\%$ of the likelihood. This quantity can be 
expressed as a fraction of the sky or in square degrees (since the whole sky is 
$4\pi, \mathrm{sr.} \approx 4.1\times10^4\, $deg$^2$).

\subsection{Discrete data} \label{sec:discrete_data}

One draw-back of the null-stream construction is that it requires the ability to take linear combinations of the data at a particular time or frequency.
In practice, PTA residuals are not observed at the same time for each pulsar, so an interpolation in time or frequency is required to use this method on real data. In the following, we make the simplifying assumption that we can work with Fourier transformed quantities $\tilde\hp$ and $\tilde\hc$.

Any \PTA\ observations will be discrete in time, and so is our simulated data. As a simplification, the simulation has $n$ data points evenly spaced in time, with cadence $\Dt$. This allows for calculating the discrete Fourier transform efficiently with the Fast Fourier Transform (FFT) algorithm \citep{CooleyTukey:1965}.
In the case of unevenly sampled data, interpolation methods can be used to estimate residuals at evenly sampled timestamps, allowing the calculation of the Fourier transform. This was addressed, for example, by \citet{2015MNRAS.449.1650Z}), who used linear interpolation between datapoints. Alternatively, Fourier coefficients for an arbitrary basis of frequencies can be directly estimated via a likelihood calculaiton for any type of data, as demonstrated in \citet{2013PhRvD..87j4021L}.

For the study of sky localisation with the null stream method, the assumption is made that a source has been detected at a known frequency $f_0$. Therefore, the likelihood calculation can be restricted to the Fourier component at this frequency. To speed up the calculation, the number of points and cadence is matched such that there is only one non-zero Fourier component. This is effected when $\Dt$ is a multiple of $(n f_0)^{-1} $, in which case bin number $f_0/\Delta f = f_0 n \Dt$ completely contains the signal power. In general, the power in a discrete Fourier transformed is spread over multiple bins and can still be recovered.

The model for the \GW\ polarizations from Eqs. \ref{eq:int_polarization_Ap} and 
\ref{eq:int_polarization_Ac} needs to be adapted to Fourier-transformed 
discrete data. The transform of the sine and cosine functions are delta 
functions, which yield a contribution at $f = -f_0$ and $f = f_0$ when 
integrated over frequency. The first can be disregarded since we only have 
positive frequencies. In the discrete transform, this power will end up spread 
over the bin corresponding to $f_0$, and so there is an additional factor 
$1/\Delta f = T$. The model then, is:
\begin{align}
\tilde{\Ap}_{(t)} (f_0) &\approx T \frac{A}{4\omega_0} (1 + \cos^2{\iota}) 
e^{i(3\pi/2 + \phi)} \\
\tilde{\Ac}_{(t)} (f_0) &\approx T \frac{A}{2\omega_0} (\cos{\iota}) e^{i(\pi + \phi)}.
\end{align} \\

The full model is $\bm{h} = (\HpF, \HcF, 0, \hdots, 0)$ (see Eq. 
\ref{eq:strainspace_data}).
As such, the model can 
be written as $\bm{h} = \bm{h}_0 e^{i\phi}$, which means that the likelihood 
from Eq. \ref{eq:likelihood} can be analytically marginalised over the phase 
$\phi$ (from $0$ to $2\pi$) (e.g. \citet{Jaranowski:2010rn}). Without explicitly writing the normalisation, the 
marginalised likelihood is given by:
\begin{align}
&p(\Mmat\bm{d}|\boldsymbol{\hat{\varOmega}}, \bm{\lambda}, \psi, \mathcal{H}_{sig}) = \int_0^{2\pi} d\phi\, p(\phi| \mathcal{H}_{sig})\, p(\Mmat\bm{d}|\phi, \boldsymbol{\hat{\varOmega}}, \bm{\lambda}, \psi, \mathcal{H}_{sig}) \nonumber\\
  &\propto \exp{(-\frac{1}{2}(|\Mmat\bm{d}|^2 + |\bm{h}_0|^2))}\, I_0(|\Mmat\bm{d}\cdot{\bm{h}_0}|).
  \label{eq:final_like}
\end{align}
Here, the dot product and norm is analogous to the product in Eq. \ref{eq:likelihood}, which is weighted by the transformed inverse covariance matrix: $a\cdot b = a^{\top}((\Mmat^{-1})^{\top} \Gamma \Mmat^{-1}) b$. The last term in Eq. \ref{eq:final_like} is the modified Bessel function of the first kind $I_0$. 

The other parameters of the SMBHB $A$, $\psi$ and $\iota$
can not be marginalised analytically. To get the likelihood for a specific sky 
location, $p(\Mmat \bm{d} | \boldsymbol{\hat{\varOmega}}, \mathcal{H}_{sig})$, the other 
parameters are marginalised numerically. This is a three dimensional integral 
over prior ranges $0-\pi$ for $\psi$ and  $0-10^{-12}$ for $A$. The prior for 
the inclination is flat in $\cos{\iota}$, with a range $-1$ to $1$.

\section{Results: sky localisation performance}
\label{sec:results}

To investigate the performance of our localisation method, we ran a set of 
simulations in which a \GW\ signal according to the SMBHB model (Eqs. 
\ref{eq:polarization_Ap} and \ref{eq:polarization_Ac}) is added to white 
noise. The likelihood as in Eq. \ref{eq:likelihood} is then calculated over a 
grid of sky locations, and marginalised over the model parameters, to determine 
$\loc$. This grid consists of 12\,288 equal area pixels made using the HEALpix 
algorithm \citep{HEALPix} via healpy\footnote{healpy.readthedocs.io}.

Simulations were carried out with a varying number of pulsars $N$ in the PTA, 
and over a range of \acp{SNR}. For $N$, the values 3, 5, 10, 20, 30, 50 and 100 
were chosen. To construct at least one null stream, 3 is the minimum number of 
pulsars needed, whereas 50 is about the number of pulsars in the combined data 
set of the current \PTA\ observatories, the International Pulsar Timing Array 
(IPTA) \citep{IPTADR1:2016}. However, IPTA pulsars are not all equally good 
timers and most of the information is carried by the ${\sim} 10$ best ones. In 
this respect, an array with $N = 50$ good timers is more comparable to what 
might be achieved in the future with the Square Kilometer Array
\citep[SKA, see][]{2015aska.confE..37J}. The range of  \SNR\ values used is 1 to 30.
This is the cumulative  \SNR\ in the PTA, i.e. summed over the pulsars:
\be
SNR^2 = \sum_{p=1}^N \sum_{i = 0}^{n-1} \frac{r_{i, p}^2}{\sigma_p^{2}}.
\label{eq:SNR}
\ee
Here, $\{r_{i,p}\}$ is the time series of $n$ residuals from pulsar $p$. The 
noise model consists of white noise in the residuals with rms $\sigma_p$ for 
each pulsar. All $\sigma_p$ are set to 100\,ns.\footnote{A more sophisticated noise model could be used by taking the product $r_i \Gammamat_{ij} r_j$ with an inverse covariance matrix $\Gammamat$.} To fix the  \SNR\  to a given value, the amplitude of the injected \GW\ is adjusted accordingly.

For each pair of $N$ and  \SNR\  values, 10 simulations were performed injecting a \GW\ source at $\theta = \pi/2, \phi = 0$, with a frequency of $20\, \text{nHz}$, and pulsars at randomised locations (with a uniform prior over the sky). These random choices are seeded such that for a given $N$, for each  \SNR\  the same 10 \PTA\ configurations are used. 300 data points were simulated with a cadence of $10^6\, \text{s}$, such that the data contain 6 full cycles of the \GW\ signal.

\subsection{Scaling with  \SNR\  at fixed $N$}
\label{sec:fixedN}

\begin{figure}
	\includegraphics[width=0.5\textwidth,clip=true,angle=0]{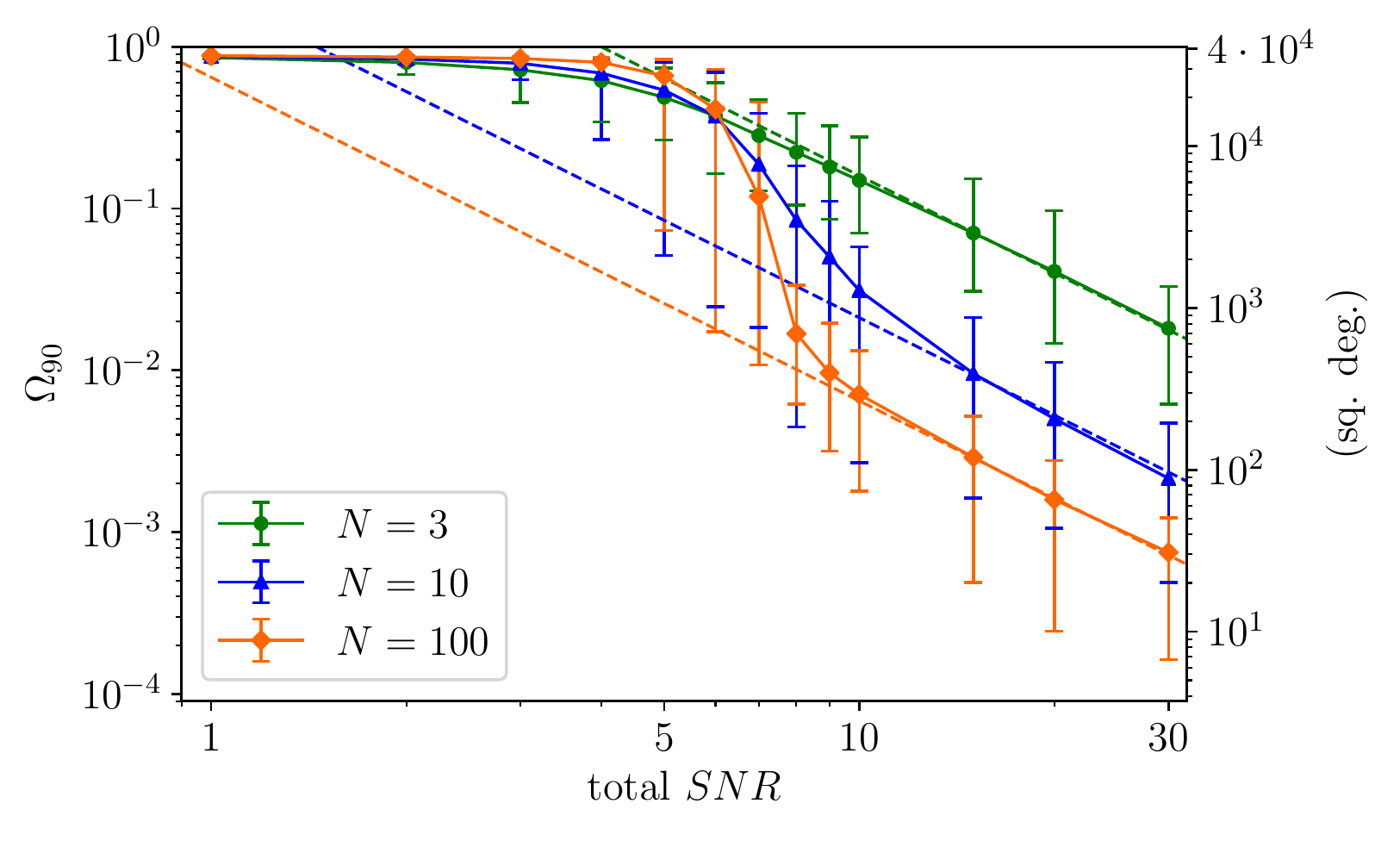}
	\caption{Fraction of the sky area containing 90\% of the likelihood 
($\loc$) vs total SNR. Data points show the mean value of 10 random realisations 
of a \PTA\ with $N =$ 3, 10 and 100 pulsars, with the errorbars showing the 
total span of results (from minimum to maximum). A power law $\loc \propto 
(SNR)^{-2}$ is fitted to the last three points of each curve. For visibility, 
not all \PTA\ sizes are plotted, but curves without errorbars are shown in Fig. 
\ref{fig:locvSNR_loglog_all}.}
	\label{fig:locvSNR_loglog}
\end{figure}

\begin{figure}
	\includegraphics[width=0.5\textwidth,clip=true,angle=0]{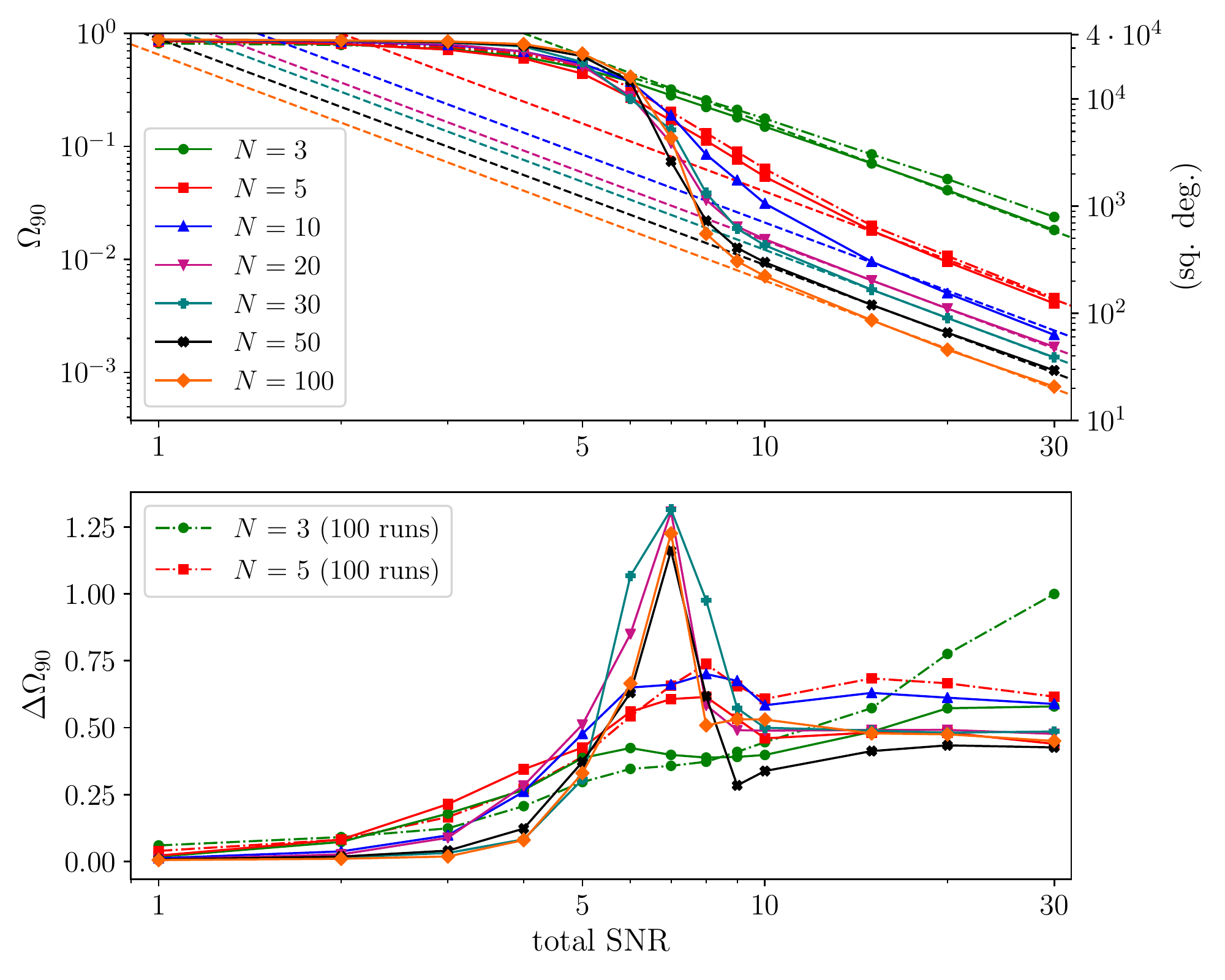}
	\caption{Top panel: Fraction of the sky area containing 90\% of the 
likelihood ($\loc$) vs total SNR, as in Fig. \ref{fig:locvSNR_loglog}. Bottom 
panel: Standard deviation normalised by the mean in the 10 realisations used for 
each point in the top panel. Dashed lines represent values as obtained with a 
re-run of 100 realisations.}
	\label{fig:locvSNR_loglog_all}
\end{figure}

We investigate the sky localisation as a function of the two main parameters identifying the detection, namely the  \SNR\  and the number of pulsars in the array.
Here we fix $N$ and vary  \SNR\  and in the following Sec. \ref{sec:fixedSNR} we 
will fix  \SNR\  and vary $N$. In Fig. \ref{fig:locvSNR_loglog}, the results for 
$N =$ 3, 10 and 100 are shown with points referring to the mean $\loc$ and 
error bars spanning the range of results in the 10 simulations for each data 
point. At low  \SNR, $\loc$ is limited to 90\% of the sky, as there is no 
information gained from the data. For $5<SNR<10$ data become informative and the 
sky localization rapidly improves, eventually converging to a $\loc \propto 
(SNR)^{-2}$ relation at high \SNR. This has to be expected since at high \SNR\ 
the likelihood surface can be approximated by a multivariate Gaussian around the 
true value of the source parameters \citep{2008PhRvD..77d2001V}. Parameter 
determination then follow the theoretical scaling $\Delta{\lambda}\propto 
(SNR)^{-1}$. Sky localization is given by a combination of the two angle 
parameters $\theta$ and $\phi$ (or equivalently right ascension and 
declination), therefore the scaling $\loc \propto (SNR)^{-2}$ is recovered.

In the region around \SNR\ from ${\sim}5$ to ${\sim}10$, a transition occurs between the two regimes (from non-informative to informative data). In Fig. \ref{fig:locvSNR_loglog_all}, the medians of the 10 runs for all values of $N$ are plotted, along with the spread in $\loc$ in the bottom panel. The transition has a similar behaviour for all $N$, but the variance around the median value is much larger for large $N$. An explanation is that for low $N$ the sky localization is still quite poor during the transition; regardless of the pulsar configuration, $\loc$ contains a significant fraction of the sky, as shown in Fig. \ref{fig:maps_N3_SNR7} for the case $N=3$. Conversely, for large $N$ the information carried by the data in the transition region strongly depends on the specific pulsar location, as shown in Fig. \ref{fig:maps_N30_SNR7} for the case $N=30$. Here we see that when some pulsars fall close to the source, its sky location is determined to high accuracy despite the low total  \SNR\  (e.g. bottom left panel); on the other hand, when there are no pulsars located close to the line of sight to the source, sky localization is poor and $\loc$ can span as much as half of the sky (e.g. bottom right panel).

\subsection{Scaling with $N$ at fixed  \SNR\ }
\label{sec:fixedSNR}

\begin{figure}
	\includegraphics[width=0.5\textwidth,clip=true,angle=0]{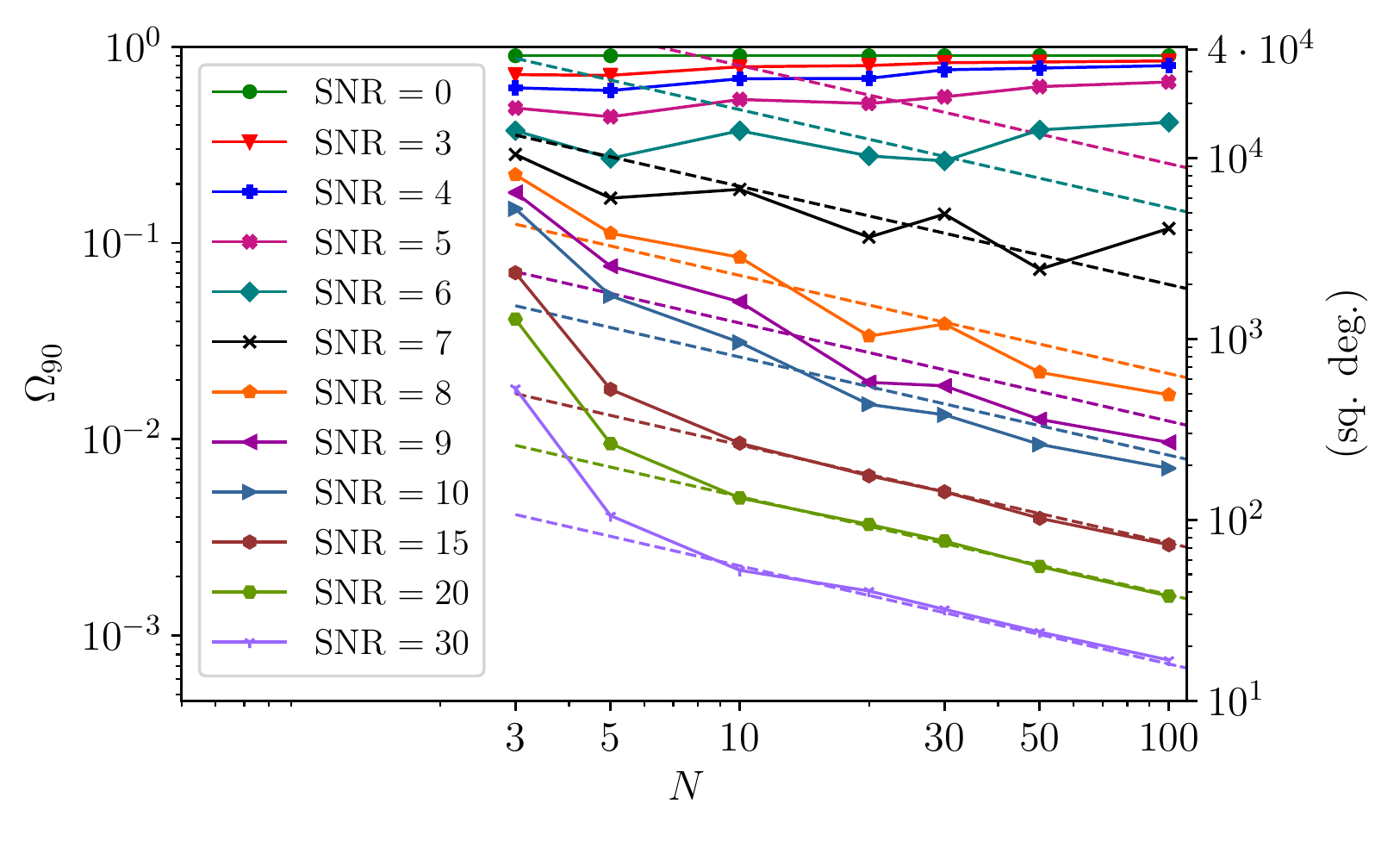}
	\caption{Fraction of the sky area containing 90\% of the likelihood ($\loc$) vs number of pulsars in the PTA. Data points show the mean value of 10 random realisations of a \PTA\ with total  \SNR\  as indicated in the inset label. A power law $\loc \propto N^{-1/2}$ is fitted to each curve, ignoring the first two points (with $N =$ 3 and 5).}
	\label{fig:locvN_loglog_all}
\end{figure}

The median $\loc$ as a function of $N$ is shown in Fig. 
\ref{fig:locvN_loglog_all} for all investigated  \SNR. For $SNR\lsim 6$ data are 
not informative and there is little dependence of the sky localisation on $N$. 
As data become informative for $SNR\gtrsim 7$, sky localisations benefits from 
increasing $N$. In the range $10\leq N\leq 100$ the improvement in sky 
localisation precision is well approximated by $\loc \propto N^{-1/2}$, 
especially for the highest  \SNR\  investigated.

A possible explanation for this scaling behaviour can be given by the average (over random PTA realizations) angular distance of the closest pulsars to the source. For increasing $N$, the angular distance between the line of sight to the source and the closest pulsar scales with $N^{-1/2}$ (for uniform randomly distributed pulsars). These closest pulsars contribute most to the sky localization (the antenna patters are modulated on the smallest scales close to the pulsar).

This conclusion is however non trivial and would need to be tested with $N > 100$. First, sky localization depends on the complex interplay of the antenna beam patterns of all the pulsars contributing to the array. Second, if the total  \SNR\  is held fix, not only the distance to the closest pulsar scales with $N^{-1/2}$, but also the  \SNR\  contributed by each individual pulsar decreases, so that the  $\loc \propto N^{-1/2}$ is not  obvious.
In any case, our systematic study indicates that for foreseeable future detections (involving a realistic number of pulsars up to 100 and  \SNR\  in the range 6-to-30) $\loc \propto N^{-1/2}$ provides a good empirical fit to the sky localization scaling.

\subsection{Dependence on source orientation}

\begin{figure}
	\includegraphics[scale=0.53,clip=true,angle=0]{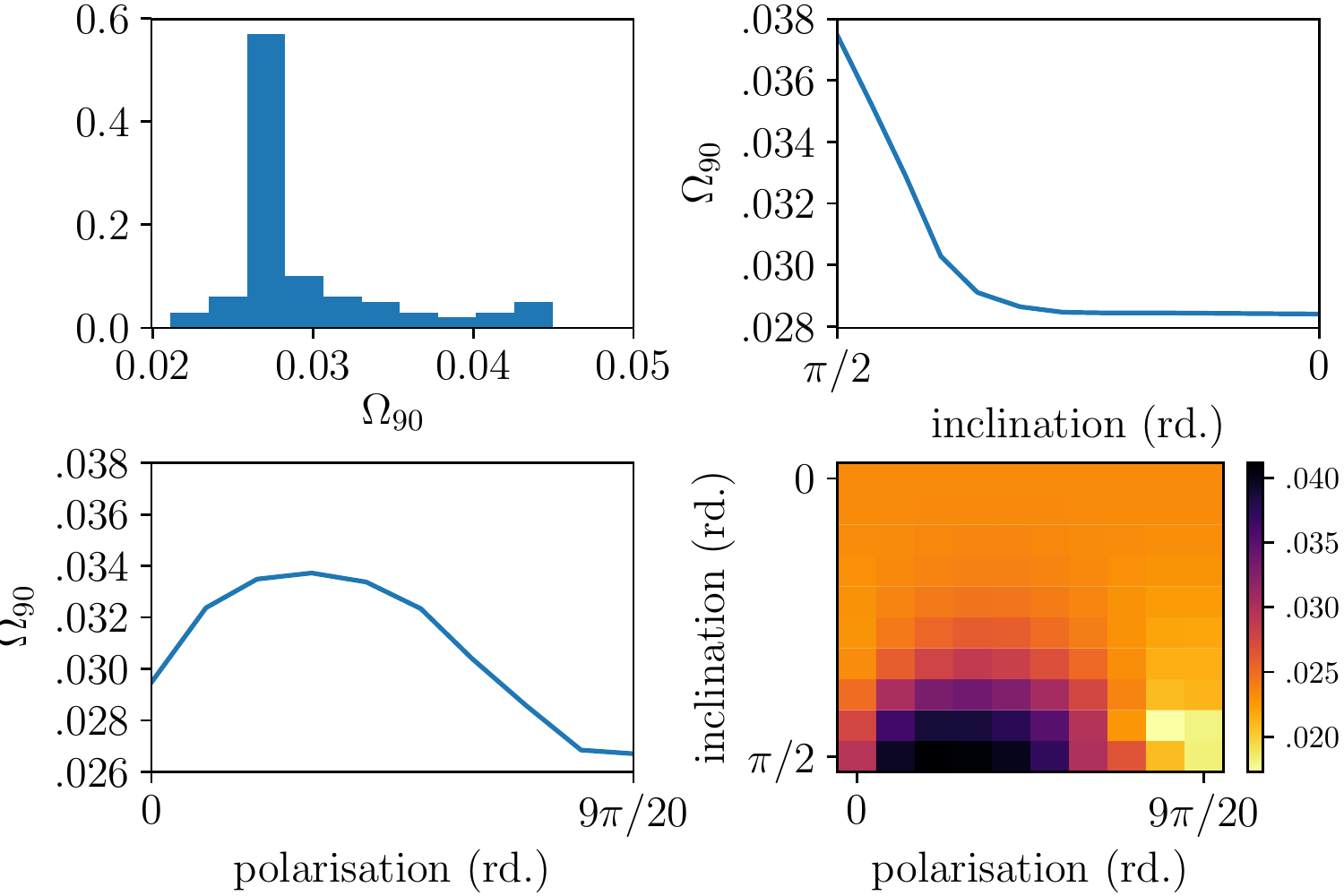}
	\caption{Distribution of sky localisations ($\loc$) obtained with varying inclinations and polarization angles of the source. Top left: Normalised histogram of $\loc$ of all 100 runs. Top right: Distribution of $\loc$ for varying inclination. Bottom left: Distribution of $\loc$ for varying polarization angle. Bottom right: All $\loc$ shown as a colour plot for the grid of polarizations and inclinations used.}
	\label{orientations_plot}
\end{figure}

So far we considered optimally oriented sources, i.e. face on systems for which the two wave polarizations equally contribute to the signal (cf Eq. (\ref{eq:polarization_Ap}) and (\ref{eq:polarization_Ac})), resulting in a circularly polarized wave. In this case the signal does not depend on the polarization angle, which only adds a contribution to the initial phase offset. Although the sky localization scalings obtained in the previous sections are expected to hold for any source inclination and polarization, the normalization of $\loc$ might depend on those quantities. To assess this dependence, we proceed as follows. We fix a \PTA\ of 10 pulsars and a source location in the sky. We then perform 100 simulations picking the source parameters from a 10$\times$10 uniform grid in polarization (chosen from $\psi = 0 - 9\pi/20$) and inclinations (chosen from $\cos(\iota) = 1 - 0$). For this particular experiment, we used noiseless data.

The bottom right panel of Fig. \ref{orientations_plot} shows the obtained value of $\loc$ on the aforementioned grid, whereas the bottom left and top right panels show $\loc$ averaged over inclination and polarization respectively. Firstly, there is essentially no dependence of $\loc$ on $\iota$ and $\psi$ so long as the former is smaller than $\approx \pi/3$. This includes about 50\% of all binaries, assuming a uniform source orientation on the sphere. 
Secondly, the average sky localisation degrades for $\iota>\pi/3$. However, compared to the reference value of $\loc = 0.028$ for the face-on case, the worst $\iota-\psi$ combination results in $\loc = 0.046$, which is a factor 1.6 worse. The average sky localisation of all the orientations with $\iota > \pi/3$, is only a factor $1.2$ worse than the face-on case. We therefore conclude that the sky localisation figures presented in Sec. \ref{sec:fixedN} and \ref{sec:fixedSNR} are a fair representation of \PTA\ capabilities for general SMBHBs.

\begin{figure*}
	\includegraphics[width=0.85\textwidth,clip=true,angle=0]{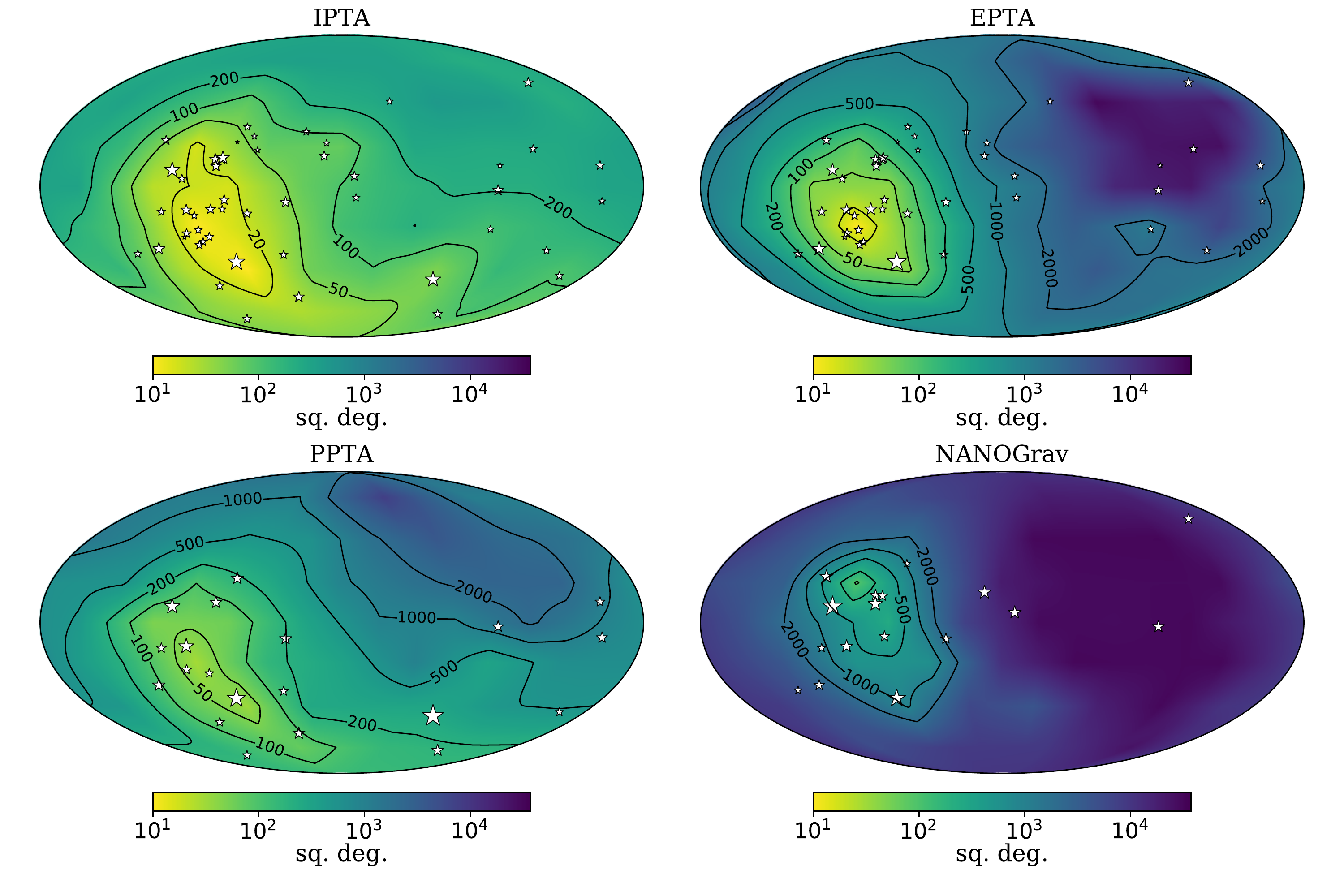}
	\caption{$\loc$ for a \GW\ source as a function of sky location for IPTA, EPTA, PPTA and NANOGrav. The simulated PTAs are approximates of the IPTA Data Release 1 and its constituent data sets: EPTA DR1, PTPA DR1 and the NANOGrav five year dataset (see text for details). The maps are interpolated from 48 pixels for which a value of $\loc$ was obtained by placing a \GW\ source with a fixed amplitude ($1.0 \times 10^{-13.5}$) in the middle of that pixel. Contours are plotted at \mbox{$\loc =$ 20, 50, 100, 200, 500 and 1000 deg\,$^2$.}}
	\label{fig:PTA_maps}
\end{figure*}

\subsection{Comparison with previous results}
Our results can be compared to previous studies dealing with systematic investigation of sky localization accuracy as a function of number of pulsars and/or \SNR\ \citep{2010PhRvD..81j4008S,Taylor:2015kpa}.
\citet{2010PhRvD..81j4008S} investigated sky localisation of individual \GW\ sources with PTAs using the Fisher Matrix formalism. Their main result is shown in their Fig. 7, where the sky localisation $\Delta\Omega$ is plotted against  \SNR\  and number of pulsars. Although results are overall compatible, there are several differences that are worth highlighting.

First, since they employ the Fisher Matrix formalism, Sesana \& Vecchio find a perfect 
$SNR^{-2}$ scaling down to $SNR=5$. Our more realistic approach shows that this 
scaling kicks in only for $SNR \gtrsim 10$, whereas for lower values, sky 
localization performances are much poorer. For example, at $SNR=5$ PTAs have 
essentially no source localisation power and even at 
$SNR=7$, typical performances are a factor of $\approx 3$ worse than the 
$SNR^{-2}$ extrapolation. This is particularly relevant since the signal builds 
up slowly with time, which means that the first confident single source \PTA\ 
detection will necessary have low \SNR. PTAs
will therefore have limited capabilities to pin down the source parameters in the early 
stages of detection.

Second, Sesana \& Vecchio found that the $N^{-1/2}$ scaling does not hold in general. Their Fig. 7 shows that the sky localisation improvement flattens out for $N>100$, even though an $N^{-1/2}$ line might provide a reasonable fit in the $10\leq N\leq 100$ range investigated in this work. It is likely that a saturation point is reached when the average contribution to the  \SNR\  of the closest pulsars of order $1$. At that point, the signal added in each pulsar (if we keep the total  \SNR\  fixed) will be below the typical noise level, and no information about the source sky localisation can be gained.

Third, the normalization of the sky localisation performance is different. For 
$N=100$ and $SNR=10$, Sesana \& Vecchio find a median $\Delta\Omega\approx 
40$deg$^2$, to be compared to our value of about $200$deg$^2$. This is partly 
due to the different definition of $\Delta\Omega$, which in their study is the 
region of the sky with probability $e^{-1}\approx 0.63$ of hosting the source. 
For a multivariate Gaussian likelihood surface, this area is a factor 2.3 
smaller than that enclosing the 90\% probability that we use. The 90\% 
probability region of Sesana \& Vecchio is therefore $\approx$100deg$^{2}$, 
which is only a factor of two smaller than what we find. Fisher Matrix calculation however, provide a lower limit to the sky 
localisation accuracy. Even for $N=100$ and $SNR=10$ we find that the 
likelihood function is highly non Gaussian, resulting in a slightly worse 
localisation performance compared to the theoretical limit.

\citet{Taylor:2015kpa} constructed a Bayesian pipeline for detection and
parameter estimation of {\it eccentric} binaries and carried out a systematic
investigation of parameter errors as a function of \SNR. Although the addition of eccentricity increases the complexity of
the problem, we do not expect this parameter to couple with the source sky
localization, and the results should be comparable with those
of our analysis.

The relevant result for comparison is reported in their Fig. 9, that
shows $\Delta\Omega$ as a function of \SNR\ for a PTA of 18 pulsars with
the properties of those used for the NANOGrav 9-year GW upper limit
\citep{2016ApJ...821...13A}. The trend of $\Delta\Omega$ with \SNR\ is
very similar to what we found, showing an initial 'transition phase' up to
about $SNR\approx 8$, then settling into the $SNR^{-2}$ behavior predicted
in the strong signal limit. The overall normalization of the curve is also
comparable. At $SNR=20$, they find a 95\% probability region ($\Omega_{95}$)
of $\approx$500deg$^{2}$, which is a factor of a few worse than the 
$\Omega_{90}$ shown in our Fig. \ref{fig:locvSNR_loglog}
for 10 and 20 pulsars, but comparable
to the 5 pulsar case. This is likely due to the fact that the 18 pulsars they use are not randomly distributed in the sky and have different noise rms, therefore only the few best contribute
significantly to the sky localization. Overall, we deem our results to be in agreement
with those of \citet{Taylor:2015kpa}.

\section{Implications for current Pulsar Timing Arrays}
\label{sec:implications}

\begin{figure*}
	\includegraphics[width=0.85\textwidth,clip=true,angle=0]{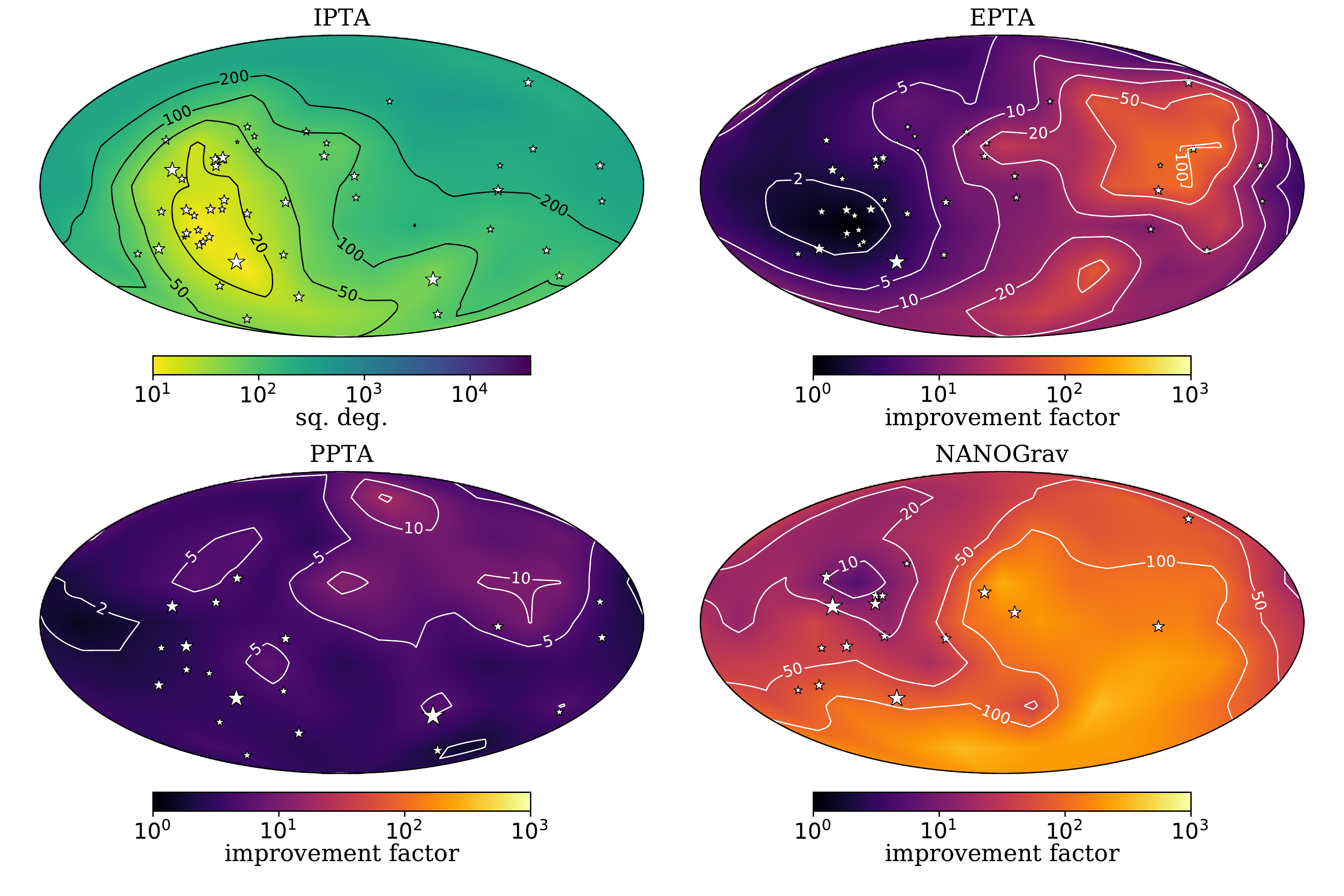}
	\caption{The top left panel is the same as in Fig. \ref{fig:PTA_maps}, whereas the remaining panels show the improvement factor ${\cal R}_X(\bm{\hat{\varOmega}})$ of IPTA compared to EPTA, PPTA and NANOGrav (see definition in the main text). The maps are interpolated from 48 pixels, as in Fig. \ref{fig:PTA_maps}. Contours are plotted at ${\cal R}_X(\bm{\hat{\varOmega}})=$ 2, 5, 10, 20, 50 and 100.}
	\label{fig:PTA_ratiomaps}
\end{figure*}

The null stream formalism developed in this work can be used to assess sky localisation capabilities of current PTAs. In the previous section, we demonstrated the beneficial effect on sky localisation of higher  \SNR\  and larger number of pulsars in the array. The obvious way to increase  \SNR\  and number of pulsars is to combine individual \PTA\ datasets under the umbrella of IPTA. In this section we therefore focus on the potential gain of IPTA for individual source localisation.

With the aforementioned goal in mind, we need to compare the capabilities of an IPTA dataset to those of the individual \PTA\ data that went into its production. The only official IPTA data release to date is IPTA DR1, presented in \cite{IPTADR1:2016}. We therefore use:
\begin{itemize}
\item EPTA Data Release 1, presented by \cite{2016MNRAS.458.3341D}, consisting of 42 MSPs monitored with radio telescopes at  Effelsberg, Jodrell Bank, Nancay and Westerbork;
\item the extended PPTA Data Release 1, presented by \cite{2013PASA...30...17M}, consisting of 20 MSPs monitored with the Parkes radio telescope;
\item NANOGrav five year dataset, presented by \cite{2013ApJ...762...94D}, consisting of 17 MSPs, monitored with the Arecibo and Green bank radio telescopes;
\item IPTA DR1, presented by \cite{IPTADR1:2016}, consisting of the combination of the three aforementioned datasets, for a total of 49 MSPs.
\end{itemize}

Several MSPs are monitored by multiple regional PTAs, and so the number of MSPs in IPTA does not correspond to the sum of those in the regional PTAs. However, by combining multiple datasets, IPTA features more high quality pulsars than the regional PTAs. We also stress that we considered the regional \PTA\ data releases that were used to build IPTA DR1, which is the meaningful thing to do since our scope is to assess the benefit of combining \PTA\ data. 

\begin{figure*}
	\includegraphics[width=0.85\textwidth,clip=true,angle=0]{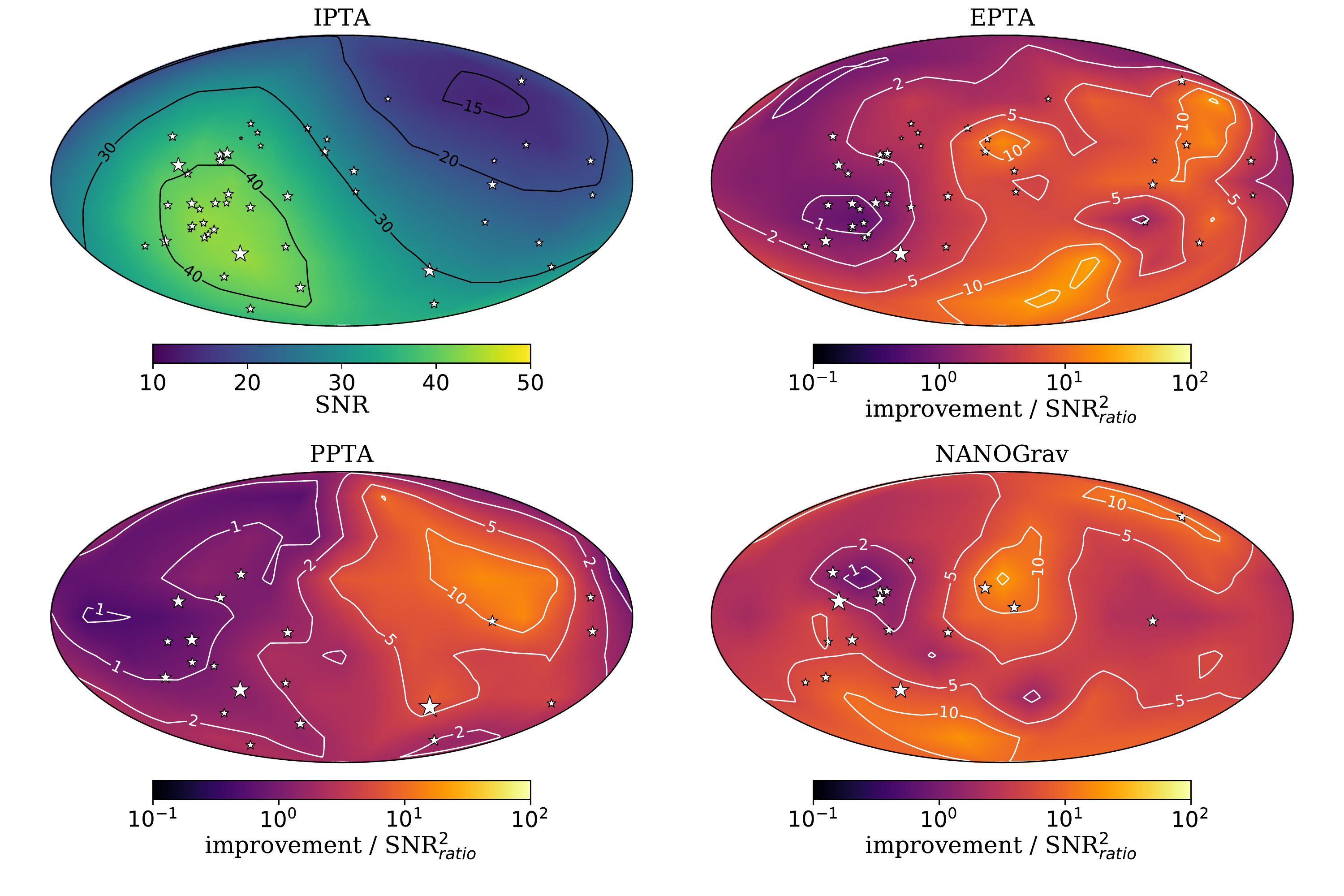}
	\caption{The top left panel shows the source  \SNR\  as a function of sky location in the IPTA. The remaining plot show the ' \SNR\  gain normalized' improvement factor $\tilde{\cal R}_X(\bm{\hat{\varOmega}})$ of IPTA compared to EPTA, PPTA and NANOGrav (see definition in the main text). The maps are interpolated from 48 pixels, as in Fig. \ref{fig:PTA_maps}. Contours are plotted at a value of $\tilde{\cal R}_X(\bm{\hat{\varOmega}})=$ 1, 2, 5, 10 and 20.}
	\label{fig:PTA_improvementvSNR}
\end{figure*}

The current implementation of our technique allows to use different rms residuals and arbitrary sky location for each individual pulsar, but is only applicable to evenly sampled data spanning the same observation time. We therefore need to modify the \PTA\ data accordingly, while keeping their properties as close as possible to the originals. For each \PTA\ we therefore compute an average dataset length $\bar{T}$ as
\begin{equation}
  \bar{T}=\frac{1}{N}\sum_{p=1}^N T_p,
\end{equation}
where the index $p$ runs over all pulsars in the array and $T_p$ is the dataset length of the $p$-th pulsar. Likewise, we compute an average number of TOAs per pulsar $\bar{n}$ as
\begin{equation}
  \bar{n}=\frac{1}{N}\sum_{p=1}^N n_p,
\end{equation}
where $n_p$ is the number of TOAs of the $p$-th pulsar in the array. We then round $\bar{n}$ to the next integer. The values of $\bar{T}$ and $\bar{n}$ for each \PTA\ are given in Table \ref{tab1}. Individual pulsar residual rms values are used as reported in \cite{2016MNRAS.458.3341D} (their Table 1 under rms) for EPTA, in \cite{2013PASA...30...17M} (their Table 7 under Rms res.) for PTPA, in \cite{2013ApJ...762...94D} (their Table 2 under rms) for NANOGrav, and in \cite{IPTADR1:2016} (their Table 4 under Residual rms) for IPTA.

\begin{table}
\begin{center}
\begin{tabular}{c|ccc|cccc}
PTA & $N$ & $\bar{n}$ & $\bar{T}$[yr] &  \SNR\  & $\loc$[deg$^2$]& ${\cal R}$ & $\tilde{\cal R}$\\
\hline
EPTA & 42 & 592 & 12.7 & 19.4 & 4492 & 22.0 & 3.3\\
PPTA & 20 & 186 & 6.3 & 21.8 & 949 & 5.0 & 2.2\\
NANOGrav & 17 & 50 & 4.8 & 8.0 & 14172 & 102.2 & 18.8\\
IPTA & 49 & 1401 & 11.1 & 28.7 & 167 & -- & -- \\

\end{tabular}
\caption{For each PTA, we list the number of pulsars $N$, the average number of TOAs per pulsar $\bar{n}$ and the average dataset length $\bar{T}$. Also listed are the performances of each \PTA\ for a face on source with $A=10^{-13.5}$ averaged over position in the sky:  \SNR, sky localization $\loc$, and improvement factors ${\cal R}$ and $\tilde{\cal R}$ of IPTA compared to regional PTAs (see text for details).}
\label{tab1}
\end{center}
\end{table}

Now that we have specified the properties of the PTAs, we conduct our experiment by considering a face-on circular SMBHB producing a monochromatic \GW\ signal with frequency $f=20$ nHz and amplitude $A=10^{-13.5}$, well within the reach of all PTAs. We place the source in turn at 48 different points in a grid over the sky and use the null stream technique described in Sec. \ref{sec:method} to compute the $\loc$ sky localisation.

Results are presented in Fig. \ref{fig:PTA_maps}, where contours have been interpolated over the grid. Firstly, the uneven pulsar distribution in the sky results in a very source position-dependent sky localisation precision. This is particularly true for EPTA and NANOGrav that have localising power mostly in the left side of the map, where all the best pulsars are concentrated, but also for PPTA and IPTA to a lesser extent. 

Secondly, the sky localisation performance differs between PTAs. Due to the limited number of good pulsars and of the short data span, the NANOGrav five year dataset performs poorly. EPTA and PPTA on the other hand have comparable capabilities, even though the latter performs better in the right half of the map. The IPTA dataset gives the best localisation overall. The \GW\ source can be localised to better than 20deg$^2$ over a region of the sky of about 3500deg$^2$ whereas a comparable precision is achieved only by EPTA, on a smaller region of $<1000$deg$^2$. On the opposite end, IPTA can locate the source to better than 500deg$^2$ regardless of its sky location and to better than 200deg$^2$ over two thirds of the sky. For comparison, PPTA can locate the source to better than 500deg$^2$ in about half of the sky, and in some regions localisation is worse than 2000deg$^2$. On average IPTA can localise the source within 167deg$^2$ whereas EPTA PPTA and NANOGrav can localise the source within 4492deg$^2$ 949deg$^2$ and 14172deg$^2$, respectively (cf Table \ref{tab1}).

We can then define a relative improvement factor of IPTA sky localisation with respect to regional PTAs as a function of the source location $\bm{\hat{\varOmega}}$ as
\begin{equation}
{\cal R}_X(\bm{\hat{\varOmega}}) = \frac{\Omega_{90, \mathrm{IPTA}}(\bm{\hat{\varOmega}})}{\Omega_{90, X}(\bm{\hat{\varOmega}})},
\end{equation}
where $X$ stands for EPTA, PPTA or NANOGrav. This relative improvement is shown in Fig. \ref{fig:PTA_ratiomaps}. Compared to the best regional dataset (PPTA), sky localisation improves by more then a factor of two virtually everywhere in the sky, and up to a factor of ten in some regions, confirming the superior performance of IPTA.

As shown in Sec. \ref{sec:results} the sky localisation naturally improves as 
$SNR^{-2}$, but also (although to a lesser extent) as more pulsars are added to 
the array, even when keeping the total  \SNR\  fixed. We therefore investigate whether the benefits of the combined IPTA datasets go 
beyond the expected  \SNR\  scaling. We 
define the sky dependent ' \SNR\  gain normalized' improvement factor 
$\tilde{\cal R}_X(\bm{\hat{\varOmega}})$ as
\begin{equation}
\tilde{\cal R}_X(\bm{\hat{\varOmega}}) = {\cal 
R}_X(\bm{\hat{\varOmega}})\times\left(\frac{SNR_X}{SNR_\mathrm{IPTA}}\right)^2.
\end{equation}
By normalizing ${\cal R}$ with the square of the  \SNR\  ratios, $\tilde{\cal R}$ quantifies the improvement brought by the better IPTA sky coverage. Results are shown in Fig. \ref{fig:PTA_improvementvSNR} and highlight that IPTA benefits indeed go beyond the source  \SNR\  increment. $\tilde{\cal R}$ is larger than unity in most of the sky for all regional PTAs. (Exceptions are a fourth, a sixteenth and a forty eighth of the sky for PPTA, EPTA and NANOGrav, respectivey. These are the areas where sky localization is best for the regional PTAs). In all cases, gain factors of up to 10 are found in parts of the sky, where the beneficial effect of better sky coverage of IPTA is maximized. Averaged over the sky, we have  $\tilde{\cal R}=3.3, 2.2, 18.8$ for EPTA, PPTA and NANOgrav respectively, certifying the benefits of IPTA data combination. We remark that the great improvements compared to NANOGrav are simply because only their five year dataset was included in IPTA DR1. An IPTA DR2, including the nine year NANOGrav dataset is currently under construction; this will allow to verify the benefits of IPTA when combining three datasets of comparable quality.

We caution that these results have been obtained by using an average timestamp for all pulsars of each specific array. In practice, PTA data are not evenly sampled and the timespan of observations varies from pulsar to pulsar. We expect, however, that considering more realistic PTAs would only have a minor impact on our conclusions. Here we consider typical resolvable sources at a frequency of several tens of nHz. So long as the cadence of observations is much shorter than the GW period, the assumption of evenly sampled data should not really matter. Furthermore, although the cadence and timespan of individual pulsars are different, they usually lie within a factor of two of the average values that we assumed in Table \ref{tab1},  again suggesting that by using the actual timestamp of each pulsar we should reach similar conclusions.  Nonetheless, it is important to verify these expectations by employing an algorithm that can handle the complexity of more realistic datasets, an extension that we plan to explore in future work.

\section{Conclusions}
\label{sec:conclusions}
In this paper, we introduced a general mathematical description for the construction of null streams in response of an individual \GW\ source. This method is general, works both in the time and frequency domain and can be applied to any deterministic waveform. We then provided a Bayesian framework to extract the \GW\ source parameters by exploring the likelihood given by the comparison of the constructed null streams and theoretical model. As proof of concept, we applied the method to the special case of a monochromatic \GW\ source generated by a circular SMBHB, considering the Earth term only in the \PTA\ response function. We used this setup to carry out a systematic investigation of \PTA\ sky localization capabilities as a function of the array parameters using the sky region containing 90\% of the source location likelihood distribution ($\Omega_{90}$) as figure of merit.

We found that for $SNR\gtrsim 10$, $\Omega_{90}$ scales as $SNR^{-2}$, as 
expected from theoretical arguments in the high  \SNR\  limit. However, we find that at low \SNR\ this scaling breaks down,
and the source cannot be well-localised. A transition between the two regimes is found for $5 \lsim SNR \lsim 10$, in which the $\Omega_{90}$ 
improvement is much steeper than the theoretical scaling. $\Omega_{90}$ is also 
found to scale as the inverse square root of the number of pulsars in the array 
$N^{-1/2}$, at least for $10 < N < 100$ and $ SNR \gtrsim 8$. As a reference point, the median 
$\Omega_{90}$ for a \GW\ source observed with $SNR=10$ in an array of 100 equal 
MSPs randomly distributed in the sky is about 200deg$^2$. These results are 
generally consistent with previous findings based on Fisher Matrix calculation, 
although there are significant differences in the $5<SNR<10$ transition region 
and in the $\Omega_{90}$ normalization.

We then used our formalism to investigate the sky localisation capabilities of regional PTAs compared to IPTA. We found that the benefits of combining data in the IPTA framework go beyond the mere gain in  \SNR\  due to the accumulation of a larger amount of data. When normalized by  \SNR\  gain, IPTA is found to perform a factor between $\sim$2 and $\sim$20 better than regional PTAs. This is because combining \PTA\ data provides a better sky coverage and increases the number of high quality pulsars that contribute informative data to the detection. These findings demonstrate that combining regional data under the IPTA umbrella maximises the scientific potential of PTAs as \GW\ detectors.
 
The framework we applied in this study can be improved in several ways
and extended to study a number of problems relevant to \PTA\ data analysis.
In particular, our current implementation requires that data are taken at
simultaneous times in all pulsars if it were to be applied to a time-domain analysis.
One of our primary future goals is to develop an implementation that can handle arbitrary datasets, with unevenly sampled data, gaps, and different time spans, thus allowing the assembly of a pipeline that can be applied to real data.

We also considered only the Earth term of the \GW\ signal which may or may not be appropriate for the loudest SMBHBs, as shown in \citet{2015MNRAS.451.2417R}. 
If the frequency of the pulsar and Earth term cannot be separated, then, while the Earth terms may still be cancelled by the null-stream method, there will remain a contribution to the power from the pulsar terms. This could be treated as an excess unmodelled (noise) power, or may be modelled explicitly by introducing an additional amplitude and phase term for each pulsar.
Efficient methods exist to either maximise or marginalize the calculation over
these parameters, as shown for example by \citet{2016MNRAS.461.1317Z} and
\citet{2014PhRvD..90j4028T}, and this is another avenue we wish
to explore.

Of great interest is also the expansion of the formalism to treat the cases of multiple deterministic sources and stochastic \GW\ backgrounds. Besides the determination of source parameters, the null stream formalism provides a powerful tool to validate candidate \GW\ signals and assess detection significance, a possibility that we want to explore in the context of Bayesian model selection. 

\section*{Acknowledgements}
AS is supported by a University Research Fellow of the Royal Society, and acknowledge the continuous support of colleagues in the EPTA. We thank A. Possenti and S. Taylor for useful comments.
JV and AV were supported by UK Science and Technology Facilities Council (STFC) grant ST/K005014/1 and ST/N000633/1, respectively. The methods for this work are implemented using the Python programming language\footnote{www.python.org}, and make extensive use of the NumPy/SciPy library \citep{NumPyArray, SciPy}. Some runs were performed with the University of Birmingham BlueBEAR HPC cluster.

\bibliographystyle{mn2e_Daly}
\bibliography{biblio.bib}

\bsp

\label{lastpage}

\appendix
\section{Selected sky maps}

\begin{figure*}
	\includegraphics[width=0.4\textwidth,clip=true,angle=0]{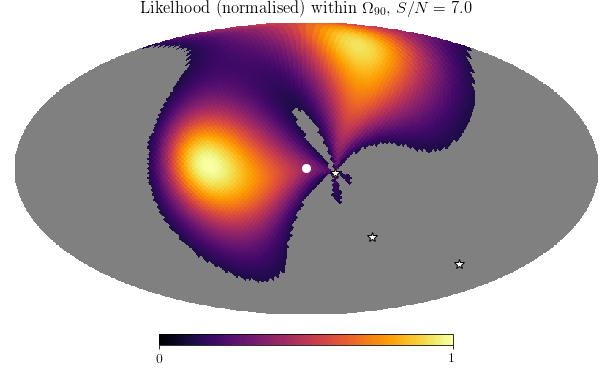}
	\includegraphics[width=0.4\textwidth,clip=true,angle=0]{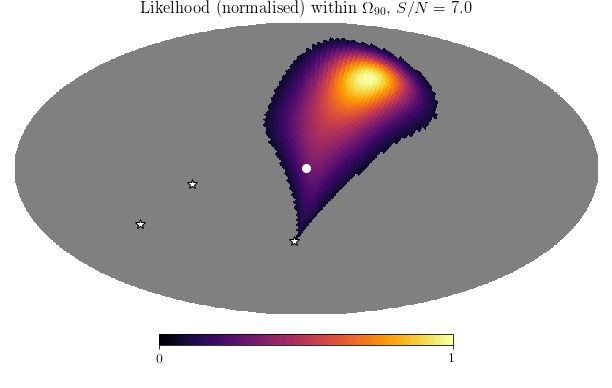}
	
	\includegraphics[width=0.4\textwidth,clip=true,angle=0]{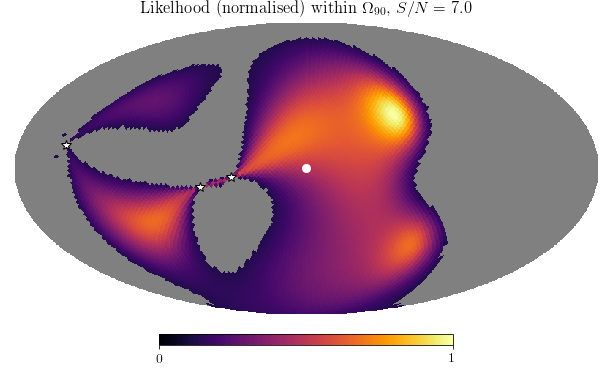}
	\includegraphics[width=0.4\textwidth,clip=true,angle=0]{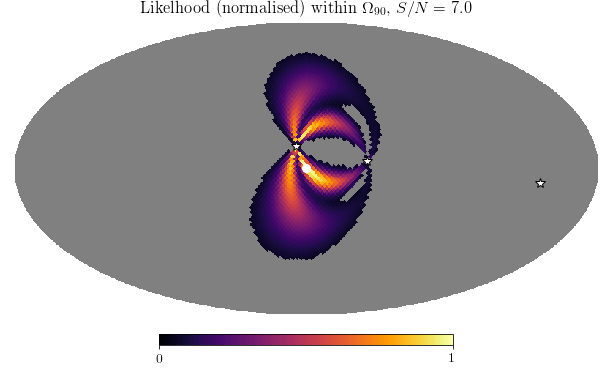}
	
	\includegraphics[width=0.4\textwidth,clip=true,angle=0]{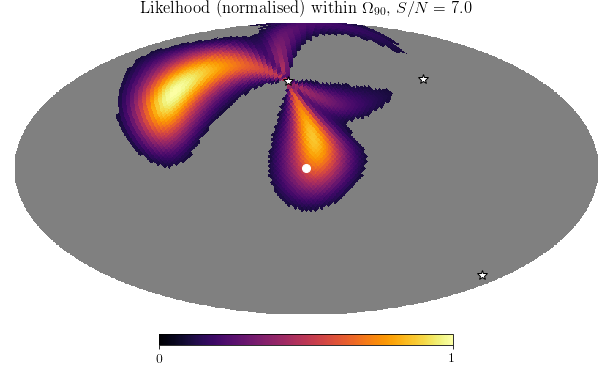}
	\includegraphics[width=0.4\textwidth,clip=true,angle=0]{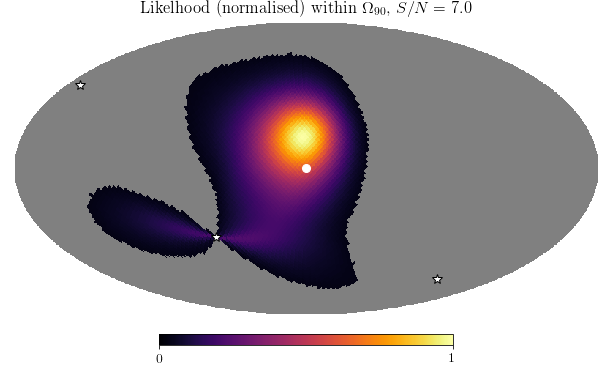}
	
	\includegraphics[width=0.4\textwidth,clip=true,angle=0]{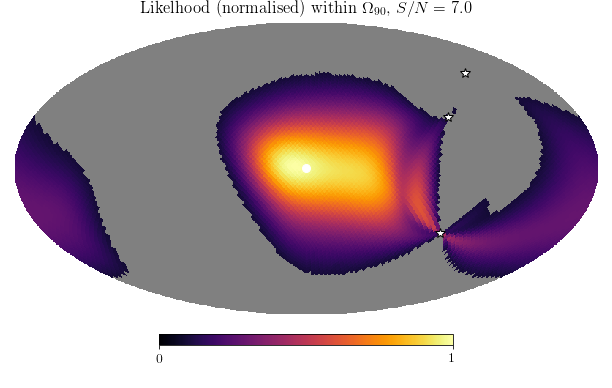}
	\includegraphics[width=0.4\textwidth,clip=true,angle=0]{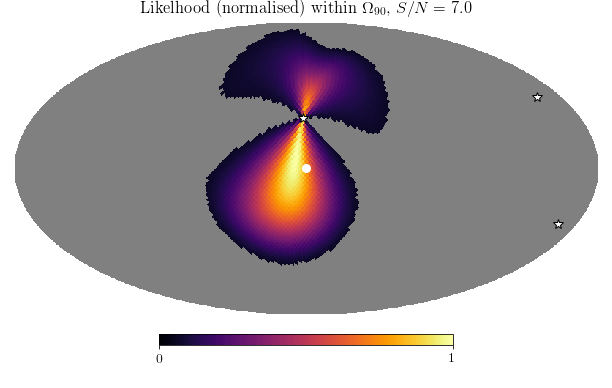}

	\includegraphics[width=0.4\textwidth,clip=true,angle=0]{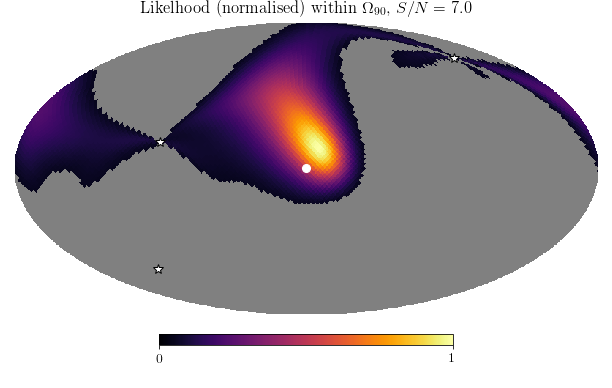}
	\includegraphics[width=0.4\textwidth,clip=true,angle=0]{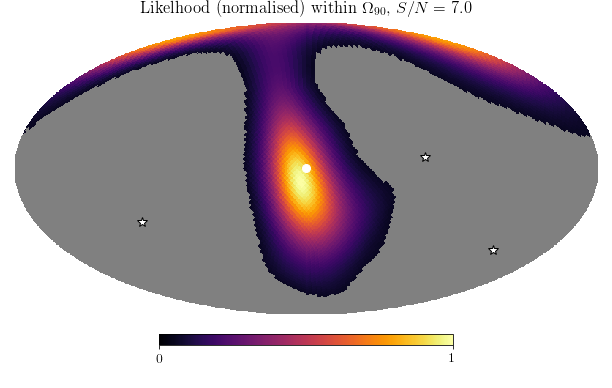}
	
	\caption{Sky maps of 10 different \PTA\ configurations with 3 pulsars, at a total  \SNR\  of 7. The injected source is always located in the middle of the map and indicated with a circle marker. The positions of the pulsars are marked with stars. Pixels not contributing to $\loc$ are masked in grey. $\loc$ ranges from 0.143 to 0.469 ($\Delta\loc = 0.563$ dex).}
	\label{fig:maps_N3_SNR7}
\end{figure*}

\begin{figure*}
	\includegraphics[width=0.4\textwidth,clip=true,angle=0]{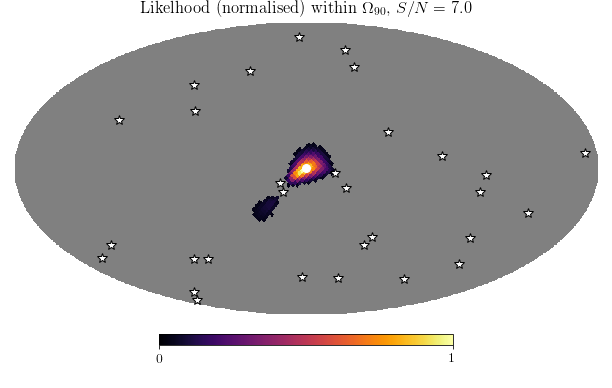}
	\includegraphics[width=0.4\textwidth,clip=true,angle=0]{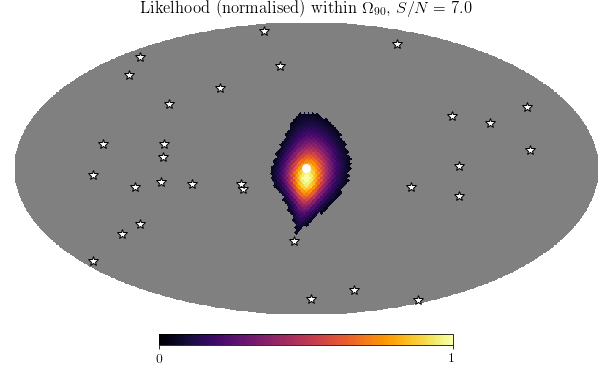}
	
	\includegraphics[width=0.4\textwidth,clip=true,angle=0]{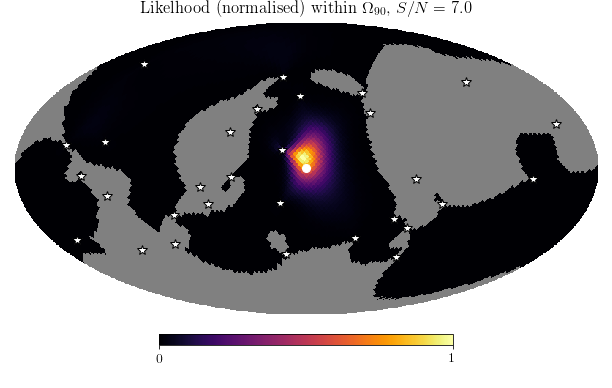}
	\includegraphics[width=0.4\textwidth,clip=true,angle=0]{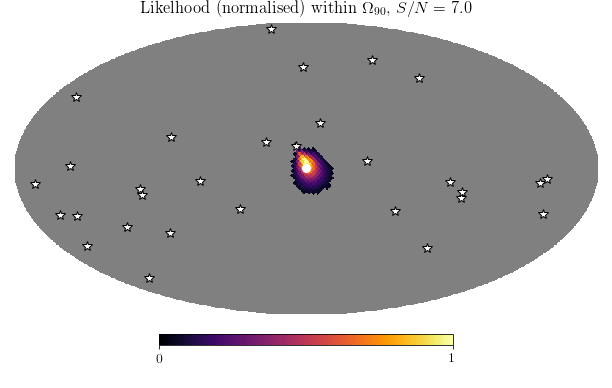}
	
	\includegraphics[width=0.4\textwidth,clip=true,angle=0]{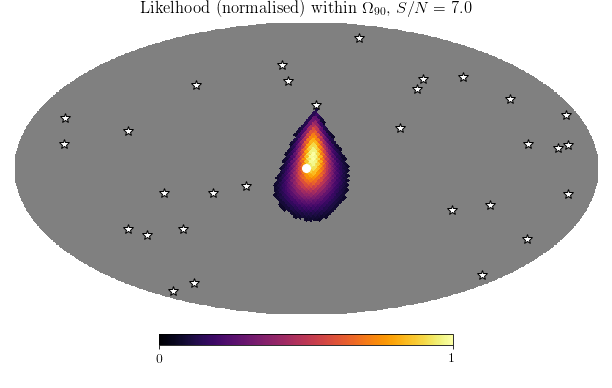}
	\includegraphics[width=0.4\textwidth,clip=true,angle=0]{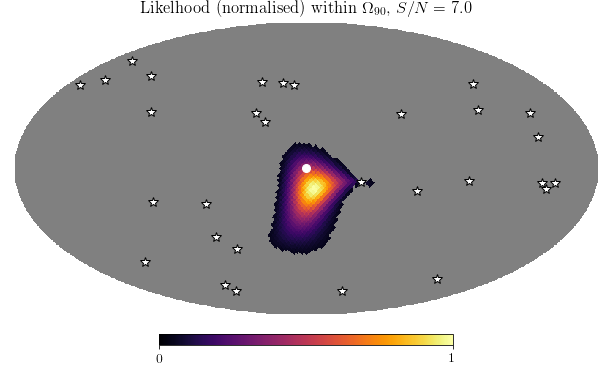}
	
	\includegraphics[width=0.4\textwidth,clip=true,angle=0]{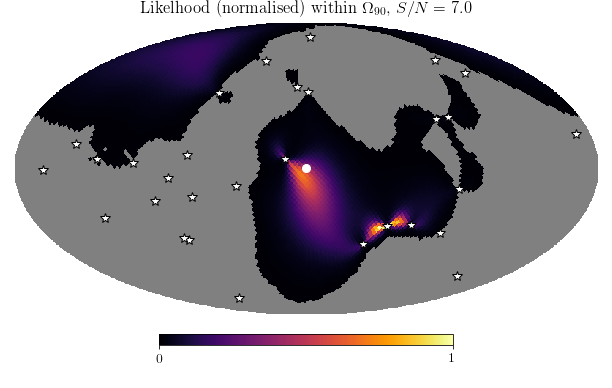}
	\includegraphics[width=0.4\textwidth,clip=true,angle=0]{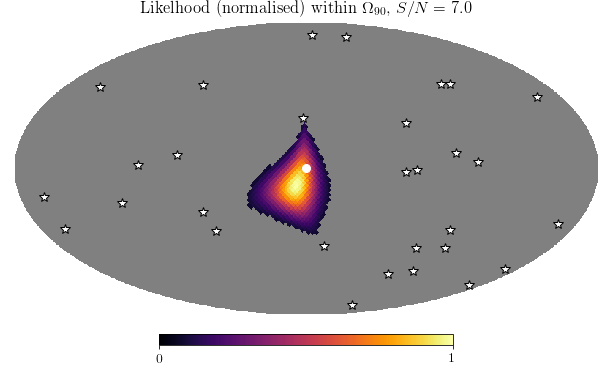}
	
	\includegraphics[width=0.4\textwidth,clip=true,angle=0]{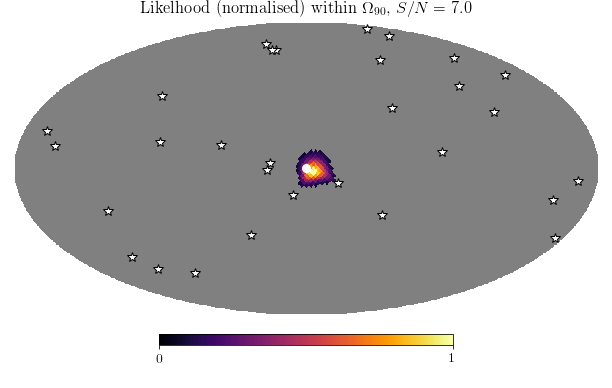}
	\includegraphics[width=0.4\textwidth,clip=true,angle=0]{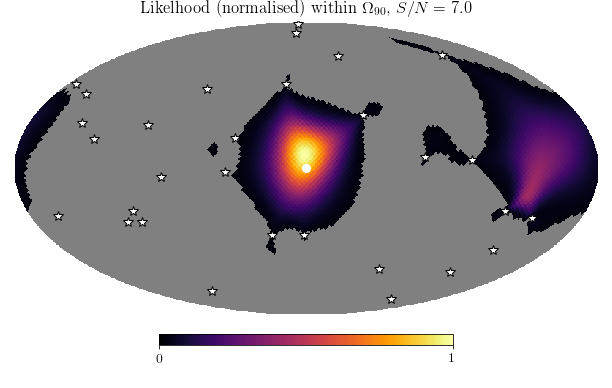}
	
	\caption{Sky maps of 10 different \PTA\ configurations with 30 pulsars, at a total  \SNR\  of 7. The injected source is always located in the middle of the map and indicated with a circle marker. The positions of the pulsars are marked with stars. Pixels not contributing to $\loc$ are masked in grey. $\loc$ ranges from 0.0083 to 0.240 ($\Delta\loc = 1.84$ dex).}
	\label{fig:maps_N30_SNR7}
\end{figure*}

\end{document}